\RequirePackage[l2tabu, orthodox]{nag}
\documentclass{stsci_report}

\usepackage{booktabs}
\usepackage{rotating}
\usepackage{natbib}
\usepackage{url}
\usepackage{graphicx}
\usepackage{caption}
\usepackage[T1]{fontenc}
\usepackage[font=footnotesize,labelfont=bf]{caption}
\usepackage{refcount} 
\usepackage[breaklinks,colorlinks=true,allcolors=blue]{hyperref}
\usepackage{color}
\usepackage{amsmath,amssymb}
\usepackage{makecell}
\usepackage[yyyymmdd]{datetime}

\usepackage[dvipsnames]{xcolor}
\definecolor{AquaGreen}{HTML}{12e193}
\definecolor{BrightSkyBlue}{HTML}{02ccfe}
\definecolor{LightRed}{HTML}{ff474c}
\definecolor{Melon}{HTML}{ff7855}
\definecolor{ShockingPink}{HTML}{fe02a2}
\definecolor{Perrywinkle}{HTML}{8f8ce7}

\newcommand{\cpspsqarc}{${\rm counts\,sec^{-1}\,arcsec^{-2}}$}
\newcommand{\arcsec}{$^{\prime\prime}$}

\renewcommand{\th}{\textsuperscript{th}}

\makeatletter 
\newcommand\footnoteref[1]{\protected@xdef\@thefnmark{\ref{#1}}\@footnotemark}
\makeatother

\copyrighttext{Copyright \copyright \the\year\ The Association of Universities for Research in Astronomy, Inc. All Rights Reserved.}

\title{\textbf{Using 23 Years of ACS/SBC Data to Understand Backgrounds}:\\Significant Reductions in Expected Background Levels}
\presubtitle{\flushleft{\includegraphics[width=5cm]{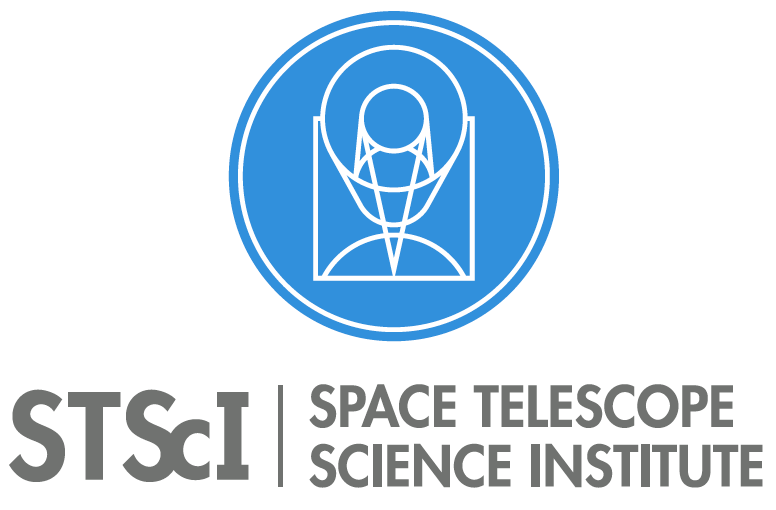}} \\ \hfill Instrument Science Report ACS 2025-04}
\author{Christopher. J. R. Clark, Roberto J. Avila, Alyssa M. Guzman, Norman A. Grogin}
\date{2025-12-30}

\begin{document}

\maketitle

\abstract{We have used 23 years of ACS/SBC data to study what background levels are encountered in practice and how much they vary. The backgrounds vary considerably, with F115LP, F122M, F125LP, PR110L, and PR130L all showing over an order of magnitude of variation in background between observations, apparently due to changes in airglow. The F150LP and F165LP filters, which are dominated by dark rate, not airglow, exhibit a far smaller variation in backgrounds. For the filters where the background is generally dominated by airglow, the backgrounds measured from the data are significantly lower than what the ETC predicts (as of ETC v33.2). The ETC predictions for `average' airglow are greater than the median of our measured background values by factors of 2.51, 2.64, 105, and  3.64, for F115LP, F122M, F125LP, and F140LP, respectively. A preliminary analysis suggests this could be due to certain O\textit{\textsc{i}} airglow lines usually being fainter than expected by the ETC. \textit{\textbf{With reduced reduced background levels, the shorter-wavelength SBC filters can conduct background-limited observations much more rapidly than had previously been expected}}. As of ETC v34.1, a new option will be included for SBC calculations, allowing users to employ empirical background percentiles to estimate required exposure times.}

\section{Introduction and Motivation} \label{Section:Introduction}

The Solar Blind Channel (SBC; \citealp{Tran2003B}) of the Advanced Camera for Surveys (ACS; \citealp{Clampin2000B}) on the Hubble Space Telescope (HST; \citealp{Bahcall1982C}) is a Multi-Anode Microchannel Array (MAMA), a photon-counting detector used for Far-UltraViolet (FUV) observations via 6 imaging filters and 2 low-spectral-resolution prisms. Although they do exhibit dark current, MAMAs like the SBC do not produce any read noise; as a result, the instrumental/astrophysical background is the main source of noise in SBC observations. However, background levels in SBC data are known to vary considerably in different circumstances, significantly affecting the sensitivity that users will obtain in a given integration time, in a given filter. 

At present, the Exposure Time Calculator (ETC\footnote{\label{Footnote:ACS_ETC}\url{https://etc.stsci.edu/etc/input/acs/imaging/}}) partially accounts for this. Using the same background model as used for other HST instruments, the ETC allows users to choose `low', `average', or `high' background levels, for each of earthshine, airglow, and zodiacal light. The resulting background estimates can be hugely sensitive to the specific user choices. For instance, with the F115LP filter, when selecting `average' zodiacal light, airglow, and earthshine, the ETC estimates a background of 69\,\cpspsqarc. However, changing the airglow to `low', but keeping everything else the same, drops the background estimate to only 12\,\cpspsqarc\ – a factor of 5.75 difference in background. Depending on source brightness, this can mean up to a factor of  5 difference in the integration time necessary for a given observation to successfully detect its target. Moreover, a typical user will often not know under what circumstances they should expect a non-`average' zodiacal/airglow/earthshine background.

As an example, FUV airglow lines have luminosities that are strongly affected by Solar activity. The most prominent lines are Ly-$\alpha$ at 1216\,\AA, and O{\sc i} at 1304\,\AA\ and 1356\,\AA, all of which are excited by Solar photons. There is also the Lyman-Birge-Hopfield system of  N$_{2}$ lines, electronically-excited by both Solar and geocoronal electrons, emitting over the 1270--2800\,\AA\ range, and which can dominate 1300--1900\AA\  \citep{Torr1994F,Eastes2000C,Cantrall2021A}. Other weaker airglow lines tend to be minor contributors to the total airglow luminosity, such as the Schumann-Runge O$_{2}$ lines at 1750--2000\,\AA\ \citep{Hedin2009A}. The impact of airglow lines upon the SBC background for a given observation will therefore affected by Solar time, Earth limb angle, and activity variations throughout (and between) Solar cycles \citep{Waldrop2013A,Putis2018A}, and more. 

The large array of factors (both known and unknown) which can affect background levels in the SBC therefore make it hard to constrain what backgrounds are `normal' or `expected' in different circumstances. Fortunately, because of the SBC's long lifetime, we have a large archive of data, consisting of many thousands of exposures, as a resource for empirically tackling this problem. In this ISR, we describe how we used this observational archive to improve our understanding of the background levels actually encountered when using the SBC.

In a companion publication, ACS ISR 2026-01 (Clark et al., 2025), we use the background levels presented here to model what observational parameters actually drive variations in the background, thereby allowing us to predict the expected background for given observation.

This ISR is primarily concerned with imaging observations with SBC. Because spectral observations with the SBC prisms use the same detector, we measure the background levels in the same manner -- however, the consequent implications for variations in {\it spectral} sensitivity are not explored. 

\subsection{Dark Current vs Other Sources of Background} \label{Subsection:Dark_vs_Background}

The SBC dark rate is often considered separately from other sources of background -- for instance, in the ETC\hyperref[Footnote:ACS_ETC]{\textsuperscript{\getrefnumber{Footnote:ACS_ETC}}}. However, the empirical measurements of SBC background levels in this study cannot distinguish between them. There is no clean way to separate dark current from other sources of background in a given observation, because not only does the dark rate vary as a function of both position on the detector and instrument temperature -- but the strength of this temperature dependence {\it also} varies with position on the detector. 

The high-dark region, from the center to upper-right of the detector, has an especially strong temperature dependence, with the dark rate increasing sharply for temperatures above 25\,$^\circ$C \citep{Avila2017B}. In contrast, the low-dark region, towards the lower-left of the detector, has a dark rate that is mostly unaffected by temperature. This is hence the location of the \texttt{SBC-LODARK} target placement aperture \citep{Avila2018A}, which is recommended for observations long enough ($\gtrsim$2\,hrs) that the detector will warm sufficiently to drive increased dark rate. This aligns with the objectives of this study, because various observational parameters that can drive the SBC dark rate are also likely to impact other sources of background. For instance, proximity to the South Atlantic Anomaly (SAA) can drive up dark rates \citep{Avila2017B} -- but SAA proximity can also increase the luminosity of atmospheric airglow lines \citep{He2020B}, which will further drive up the observed background. Even the small temperature effect from Earth being at perihelion versus aphelion is known to cause changes to the dark rate for the MAMA aboard the HST/COS \citep{Johnson2024C}.

We therefore make no attempt to systematically separate the contribution of the dark current to the observed background, versus other sources. We do, however, discuss in Section~\ref{Section:Conclusion} how the result of this study should be considered in relation to situations where the dark current is considered separately, such as the ETC.

\subsection{Report Structure} \label{Subsection:Report_Structure}

This ISR is laid out as follows:

In Section~\ref{Section:Archival_Observations}, we describe the archival SBC data we used for this study. 

In Section~\ref{Section:Background_Measurement}, we lay out our process for measuring the background levels in the SBC observations. 

In Section~\ref{Section:Discrepancy}, we discuss the discrepancies between the background levels we measure, and what is predicted by the standard ETC models. 

In Section~\ref{Section:Conclusion}, we summarize our findings, and describe the resulting considerations for users when designing SBC observations.

\section{Archive of Observations} \label{Section:Archival_Observations}

We drew upon the large archive of SBC observations, to capture the full diversity of ways in which the instrument is used. We evaluated every SBC observation in the public\footnote{Public as of 3\textsuperscript{rd} July 2024.} archive, to identify all the ones useful for our analysis. Specifically, we were interested in all of the individual {\it exposures}. 

Many SBC observations consist of an association of multiple exposures (which are then drizzled together or otherwise combined during reduction). However, the different  exposures that make up an observation, all happening at different times (often during different orbits), will be subject to different conditions that may affect the background levels. We therefore conduct our analyses on the individual exposures.

\begin{table*}
\centering
\caption{Number of suitable SBC exposures identified in each filter, for use in our background level analysis.}
\small
\label{Table:Exposure_Counts}
\begin{tabular}{lrrrrrrrr}
\toprule \toprule
\multicolumn{1}{c}{\bf Filter} &
\multicolumn{1}{c}{F115LP} &
\multicolumn{1}{c}{F122M} &
\multicolumn{1}{c}{F125LP} &
\multicolumn{1}{c}{F140LP} &
\multicolumn{1}{c}{F150LP} &
\multicolumn{1}{c}{F165LP} &
\multicolumn{1}{c}{PR110L} &
\multicolumn{1}{c}{PR130L} \\
\cmidrule(lr){1-9}
{\bf N\textsuperscript{\underline{o}}} & 261 & 321 & 804 & 1293 & 2933 & 939 & 296 & 1081 \\
\bottomrule
\end{tabular}
\end{table*}

Each exposure in the archive has a corresponding \texttt{\_raw.fits} and \texttt{\_flt.fits} file. We inspected every exposure's data in turn, to ensure they were appropriate for our analysis. Specifically, we performed the following checks on all the SBC exposures in the public archive:

\begin{enumerate}
\item We checked that each exposure contained valid data, to ensure it was not engineering or other null observations for which all pixels have a value of 0; these exposures were rejected. 
\item We rejected exposures for which the filter was was recorded as \texttt{BLOCK}, indicating the filter wheel was set to an opaque element (e.g, as part of a calibration observation, or due to bright object limits being triggered).
\item We required `jitter' data -- namely the \texttt{\_jit.fits} file -- to be available for each exposure, as in we make use of the pointing information this contains in later analysis (see ACS ISR 2026-01; Clark et al., 2025). For single-exposure observations, the \texttt{\_jit.fits} shares the same rootname (i.e., `IPPSSOOT' identifier) as the  \texttt{\_raw.fits} and \texttt{\_flt.fits}. For exposures that are part of a multi-exposure association, the \texttt{\_jit.fits} is identified in that association's \texttt{\_asn.fits} file. Some exposures do not have an associated \texttt{\_jit.fits} file present in the archive; these exposures were rejected.
\item There are long-running SBC calibration programs for monitoring dark rates and photometric stability (e.g, \citealp{Guzman2024I}); however, these are not ideally suited to constraining variations in background levels. These  programs consist of only a few observations a year and are generally conducted in the same manner each year. Whilst this maximises validity for comparing results from one year to the next, it means the backgrounds encountered only sample a relatively narrow range of observational circumstances.
\end{enumerate}

After having performed these checks, we were left with 8,640 SBC exposures suitable for our analyses. The filter-by-filter breakdown of these is given in Table~\ref{Table:Exposure_Counts}. The number of suitable exposures varies by up to a factor of 10 between filters, reflecting the relative frequency with which each filter is used.
 
\section{Background Measurements} \label{Section:Background_Measurement}

\begin{figure}
\centering
\includegraphics[width=0.975\textwidth]{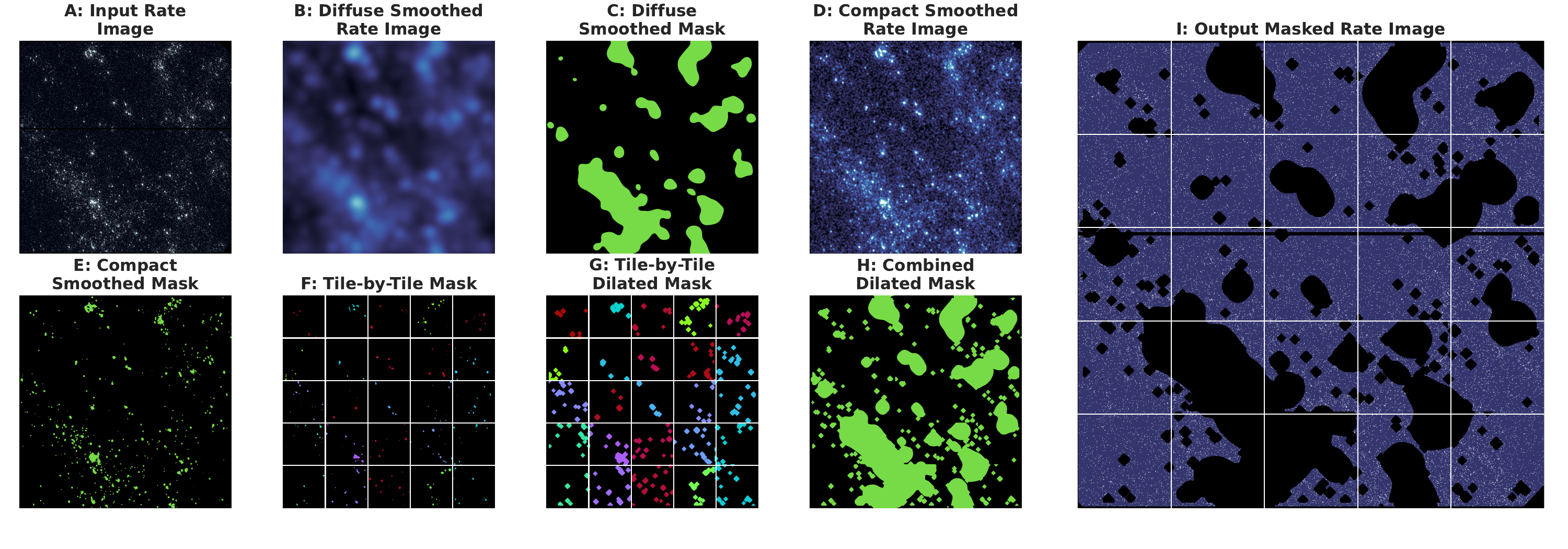}
\caption{Illustration of the steps of our masking procedure, using exposure \texttt{JEC0C6AEQ}, an F150LP observation of an area within nearby spiral galaxy NGC\,7793. Full descriptions of each step in the process are given in Section~\ref{Section:Background_Measurement}. Note that panel I, the final masked rate image, uses a different color scale than the other versions of the rate image -- panels A, B, and D -- which all share a color scale. Panel I uses a color scale tailored to the low count rates in the un-masked areas of the image, with dark blue used for pixels with a rate of 0 counts\,s$^{-1}$, and black reserved for masked regions. Panel I is also shown at twice the size of the other panels, to allow closer inspection. Panels F, G, and I show the boundaries between tiles, and the color of the masked pixels in panels F and G indicates which tile the masked pixels belong to; see Section~\ref{Subsection:Tile_Masking} for details.}
\label{Fig:Masking_Example_1}
\end{figure}

We developed a procedure for measuring the background level in SBC exposures, that was able to obtain consistent and validly cross-comparable measurements of the background for the the wide range of noise regimes and astrophysical conditions encountered in SBC imagery. The same background measuring procedure was therefore used for every exposure. 

The specific methodology for measuring the background level was developed iteratively with the methodology for using observational parameters to {\it predict} the background level, as presented in ACS ISR 2026-01 (Clark et al., 2025), the companion report to this one. As measurements of the background became more consistently precise, the models trained using those measurements became steadily more accurate. Here we describe the background-measuring method that was the result of that iterative process. Throughout, we refer to Figures~\ref{Fig:Masking_Example_1}, \ref{Fig:Masking_Example_2}, \ref{Fig:Masking_Example_3}, and \ref{Fig:Masking_Example_4}, which illustrate the process for four different exposures, each with different characteristics.

As the initial step of the background measuring process for each exposure, we first converted the \texttt{flt} frame its native units of counts, to counts\,s$^{-1}$, by  dividing by the exposure time provided in the header (under the \texttt{EXPTIME} key). We masked any pixel for which the \texttt{DQ} array had data-quality flags (most notably the bad rows, and the edge-of-field vignetting). We then applied the pixel area map corrections using the \texttt{pamutils}\footnote{\url{https://stsci-skypac.readthedocs.io/en/latest/_modules/stsci/skypac/pamutils.html}} Python library, to account for the geometric distortion of the detector

\subsection{Masking Diffuse Emission}  \label{Subsection:Diffuse_Masking}

In order to measure the background level in the exposures, we needed to mask out regions where the background is not the dominant source of counts. This requires particular care for regions with diffuse astrophysical emission, where the majority of pixels may contain no counts, and where few/no pixels individually have an elevated Signal-to-Noise ratio (S/N), but where the average count rate over an extended area is significantly elevated over the background.

An example of this would be for an observation targeting a compact region of star formation in a nearby galaxy, where lower-surface-brightness emission from the host galaxy adds significant diffuse emission over other portions of the field of view. This is exactly the situation for the exposure shown in Figure~\ref{Fig:Masking_Example_1}.

\begin{figure}
\centering
\includegraphics[width=0.975\textwidth]{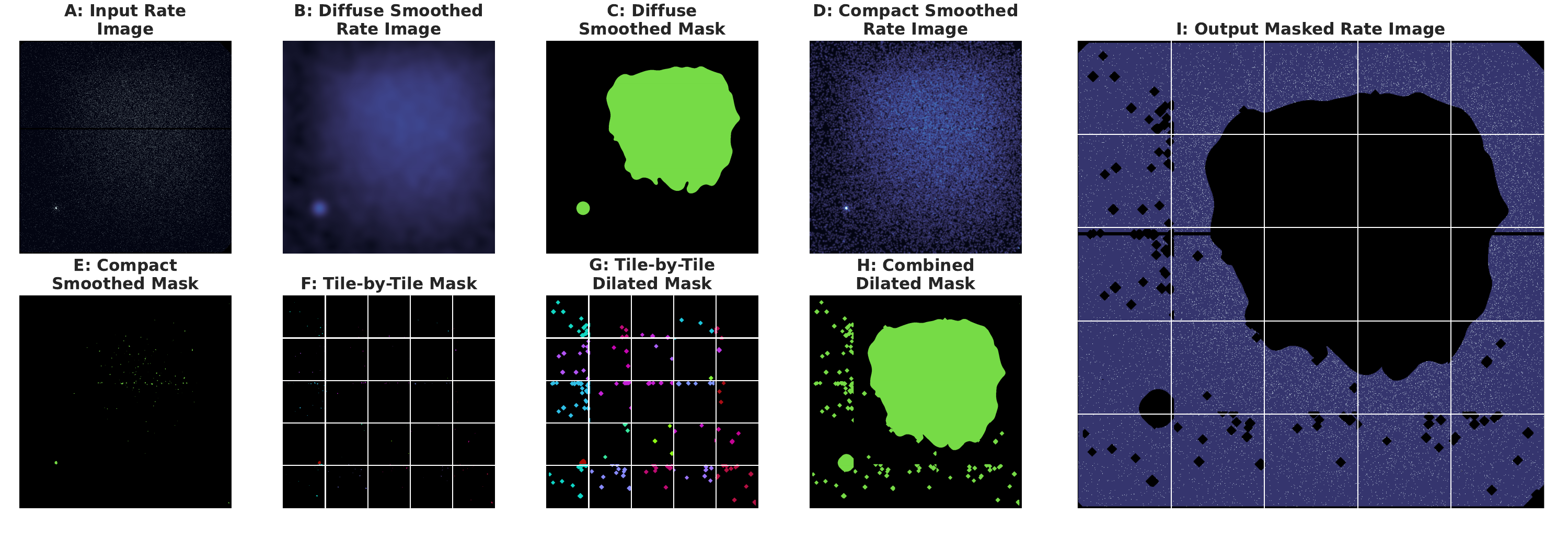}
\caption{Same as for Figure~\ref{Fig:Masking_Example_1}, except for exposure \texttt{JE2054ZIG}, an F165LP observation of globular cluster NGC\,104, which has been positioned in the low-dark region in the lower-left of the detector. Dark current is prominent in this exposure, especially in the high-dark region in the center and upper-right of the detector.}
\label{Fig:Masking_Example_2}
\end{figure}

\begin{figure}
\centering
\includegraphics[width=0.975\textwidth]{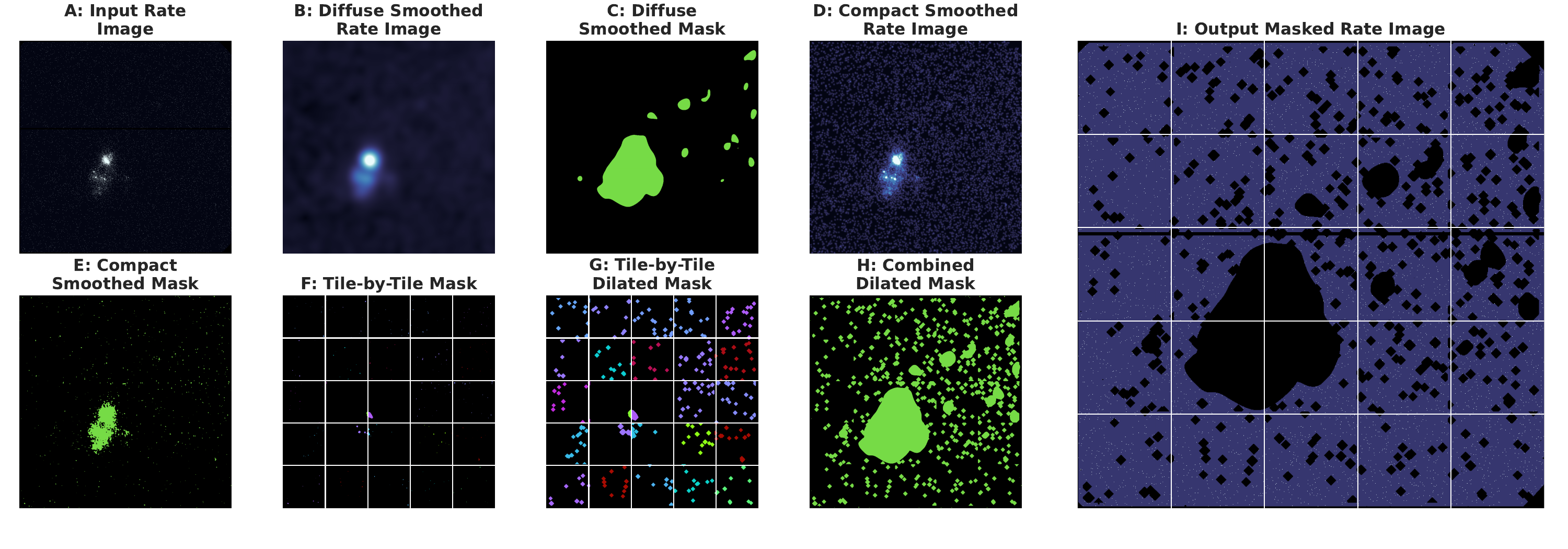}
\caption{Same as for Figure~\ref{Fig:Masking_Example_1}, except for exposure \texttt{JEDI12XPQ}, an F165LP observation of star-forming dwarf galaxy SDSS J1044+0353.}
\label{Fig:Masking_Example_3}
\end{figure}

In this situation, we need to be able to identify and mask such regions of diffuse emission, despite the individual pixels within such regions often not registering significant count rates. We do this by first smoothing the rate image by a gaussian kernel with $\sigma = 20$ pixels. This is a relatively large kernel, given the SBC detector size of only 1024$\times$1024 pixels\footnote{For reference, SBC detector pixels are approximately 0.030\arcsec\ $\times$ 0.034\arcsec, with the detector field of view therefore being roughly 30\arcsec\ $\times$ 35\arcsec.}. Smoothing by this large kernel drives up the S/N of diffuse features in the frames. This can be seen in Figures~\ref{Fig:Masking_Example_1}--\ref{Fig:Masking_Example_4}, in each of which the input rate image is shown in panel A, with panel B showing the image after being smoothed by the $\sigma = 20$ pixel Gaussian kernel.

We then perform iterative sigma-clipping of the pixel values in the frame. The first step in the sigma-clipping is finding the median, $\tilde{p}_{\it iter}$, and standard deviation, $\sigma_{\it iter}$, of the pixel values. We then define the iterative clipping coefficient, $c_{\it clip}$, such that any pixels with values that differ from $\tilde{p}_{\it iter}$ by more than $c_{\it clip}\times\sigma_{\it iter}$ are removed; for this stage in the process, we use $c_{\it clip} = 2$. This clipping is then repeated for multiple iterations, with the remaining un-clipped set of pixel values used as the input in each new iteration, until $\sigma_{\it iter}$ converges such that it differs by a factor of \textless\,0.001 between iterations (which generally takes 3--5 iterations).

Once the sigma-clipping has converged, the $\tilde{p}_{\it iter}$ and $\sigma_{\it iter}$ of the final iteration are used to define the masking threshold for the frame, such that any pixel brighter then $\tilde{p}_{\it iter} + (c_{\it mask}\times\sigma_{\it iter})$ is masked, where we use $c_{\it mask} = 3$. Note that $c_{\it mask}$ (the coefficient that sets the masking threshold) is {\it not} the same as $c_{\it clip}$ (the clipping coefficient, used for the iterative sigma-clipping process). Our choices of $c_{\it clip} = 2$ and $c_{\it mask} = 3$ were informed by empirical testing, and seem to be robust over a wide range of observational situations.

The pixels that exceed the masking threshold are indicated in panel C of Figures~\ref{Fig:Masking_Example_1}--\ref{Fig:Masking_Example_4}. As can be seen in the case of Figure~\ref{Fig:Masking_Example_1}, the process has successfully masked the regions of diffuse astrophysical emission. 

It should also be noted in Figure~\ref{Fig:Masking_Example_2} that this process also masks the extended feature towards the upper-right portion of the frame. This feature is caused by the instrument's dark current, which is most prominent in this known high-dark region of the SBC detector \citep{Avila2017B}. Whilst this study treats the dark current as part of the general background (see Section~\ref{Subsection:Dark_vs_Background}), the elevated background in the high-dark region is a specific known phenomenon, and it is already standard practice to avoid it as needed. It is therefore correct that it be masked, as the elevated background it causes should not be encountered by users in practice.

\subsection{Masking Compact Emission}  \label{Subsection:Compact_Masking}

Naturally, we also wish to mask compact sources of emission in all exposures. Bright sources are easy to identify and mask via simple pixel statistics. However, similar to diffuse emission, fainter compact sources can be a subtle matter. Often a compact source will manifest as a cluster of pixels, of which few/none are individually significant, but which together represent a strong detection. 

Therefore, as with diffuse emission, we smooth the rate image to account for this -- albeit with a much smaller kernel, suitable for compact sources. Specifically, we smooth the rate image with a Gaussian kernel with $\sigma = 2$ pixels (i.e., 10$\times$ smaller than the kernel we used for diffuse emission). The result of this smoothing is shown in panel D of Figures~\ref{Fig:Masking_Example_1}--\ref{Fig:Masking_Example_4}.

Then, once again following the process for diffuse emission, we used iterative sigma clipping to set a masking threshold. For masking compact sources, we found that a clipping coefficient of $c_{\it clip} =5$ and a masking coefficient of $c_{\it mask} =5$. Examples of the resulting compact source mask are shown in panel E of Figures~\ref{Fig:Masking_Example_1}--\ref{Fig:Masking_Example_4}.

Importantly, the compact sources detected and masked here will often not have been detected and masked by the process used for masking diffuse emission. This is because the large smoothing kernel we used to enhance diffuse emission will dilute the emission from sources much more compact than the kernel.

We also note that the Point Spread Function (PSF) of the SBC has very extended wings \citep{Tran2003B,Avila2016H}, which need to be masked to allow accurate background measurement, especially in fields with bright point sources. Fortunately, it appears that this compact source masking process does a good job of masking the PSF wings; e.g, see Figure~\ref{Fig:Masking_Example_4}.

\subsection{Tile-by-Tile Masking}  \label{Subsection:Tile_Masking}

\begin{figure}
\centering
\includegraphics[width=0.975\textwidth]{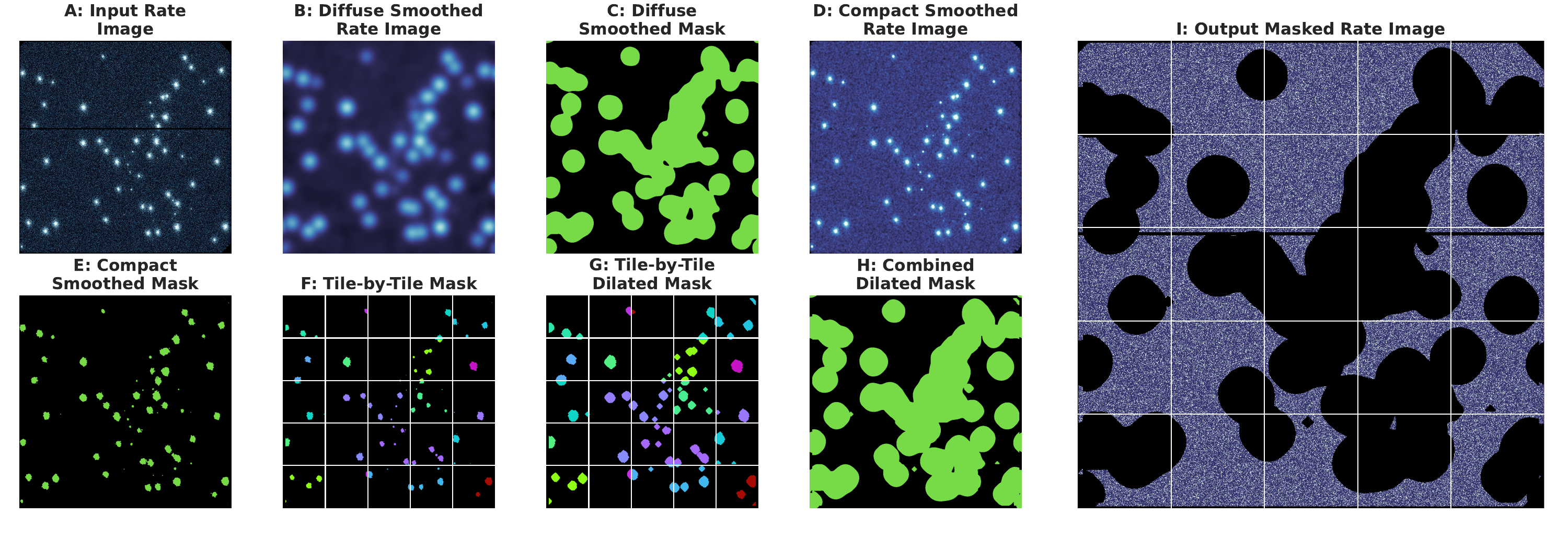}
\caption{Same as for Figure~\ref{Fig:Masking_Example_1}, except for exposure \texttt{JEG201A4Q}, an F115LP observation of open cluster NGC\,6681 (a standard SBC calibration target).}
\label{Fig:Masking_Example_4}
\end{figure}

The masking processes applied in Sections~\ref{Subsection:Diffuse_Masking} and \ref{Subsection:Compact_Masking} both compute the statistics, and apply their resulting masking thresholds, over the entire frame. However, backgrounds can vary across the frame, especially those arising from astrophysical sources (e.g, Figure~\ref{Fig:Masking_Example_1}) and dark current (e.g, Figure~\ref{Fig:Masking_Example_2}).

As a result of this, a source that would not be detectable in a high-background portion of a frame may be easily detectable in a low-background part of the same frame. However, with a masking threshold computed using the pixel value distribution of the entire frame, such a source may go undetected, hence unmasked.

We therefore perform a third round of detecting and masking regions of bright emission in each frame to address this. For this, we split the frame into a 5$\times$5 grid of 25 square tiles, each 200$\times$200 pixels in size\footnote{The SBC detector is 1024$\times$1024 pixels in size, so we therefore exclude a border of 12 pixels on each edge; a large fraction of this border is already masked in the DQ array, to account for vignetting.}. We then repeated the iterative sigma-clipping and masking within each tile, using the same slightly-smoothed version of the rate image used above in Section~\ref{Subsection:Compact_Masking} (i.e., panel D of Figures~\ref{Fig:Masking_Example_1}--\ref{Fig:Masking_Example_4}). Within each tile, we used a clipping coefficient of $c_{\it clip} =5$, and a masking coefficient of $c_{\it mask} =5$. 

By applying the iterative sigma-clipping and masking within each of the 25 tiles, we thereby apply local masking thresholds dictated by the local background properties. This maximizes our ability to detect and mask sources that would not be masked using the global approaches of Sections~\ref{Subsection:Diffuse_Masking} and \ref{Subsection:Compact_Masking}. Similar approaches are used by \texttt{photutils} \citep{Bradley2025B} and \texttt{S(ource)Extractor} \citep{Bertin1996A} to account for spatially-varying backgrounds and noise regimes.

The result of this is illustrated in panel F of Figures~\ref{Fig:Masking_Example_1}--\ref{Fig:Masking_Example_4}. The coloring of the masks indicates which of the 25 tiles the masked pixels belong to.

\subsection{Mask Dilation}  \label{Subsection:Mask Dilation}

Even with the three rounds of masking described above, we found that the regions immediately around the edges of the masks tended to have elevated pixel values, versus the rest of the frame. This isn't surprising, and represents the pixels to the periphery of regions of emission, the values of which just fell short of the masking thresholds.

To mask as many of these peripheral pixels as possible, we performed 10 iterations of binary dilation upon our masks. For each round, the mask was expanded to include any un-masked pixels adjacent to a masked pixel, thereby expanding the masked region by 1 pixel each time. Applying 10 rounds of this therefore expands the masked region by an amount corresponding to a few times the width of the instrumental PSF.

\begin{figure}
\centering
\includegraphics[width=0.475\textwidth]{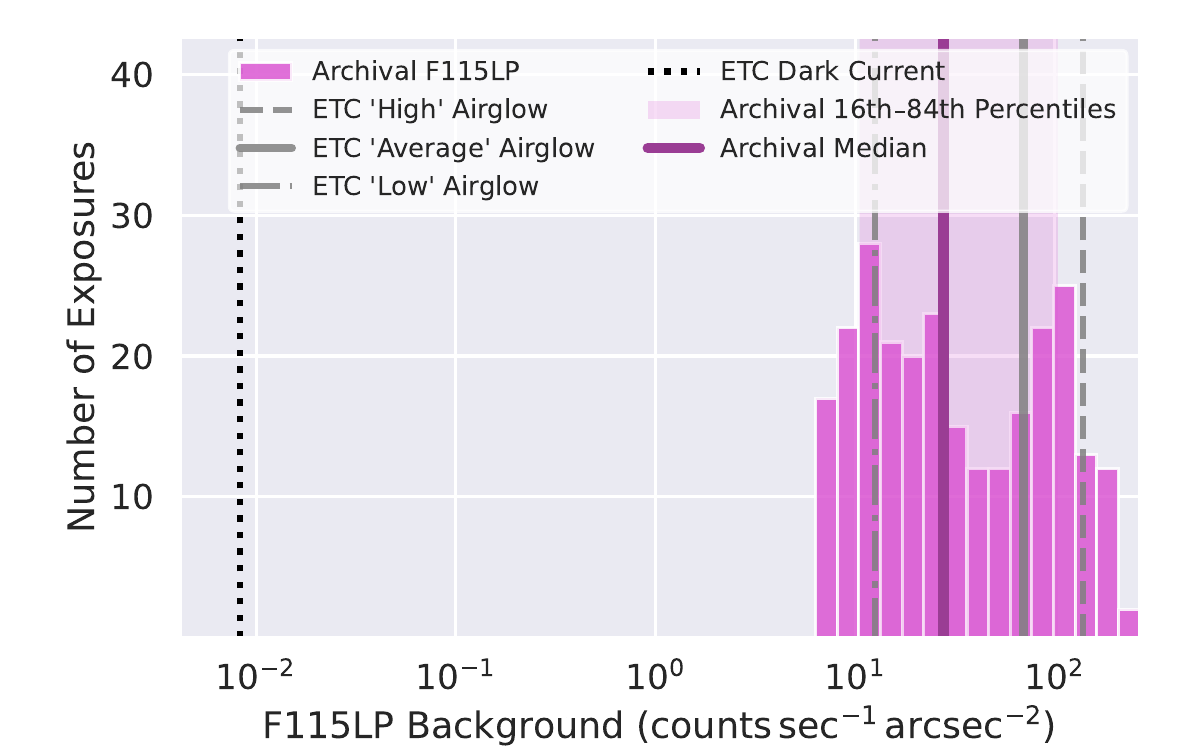}
\includegraphics[width=0.475\textwidth]{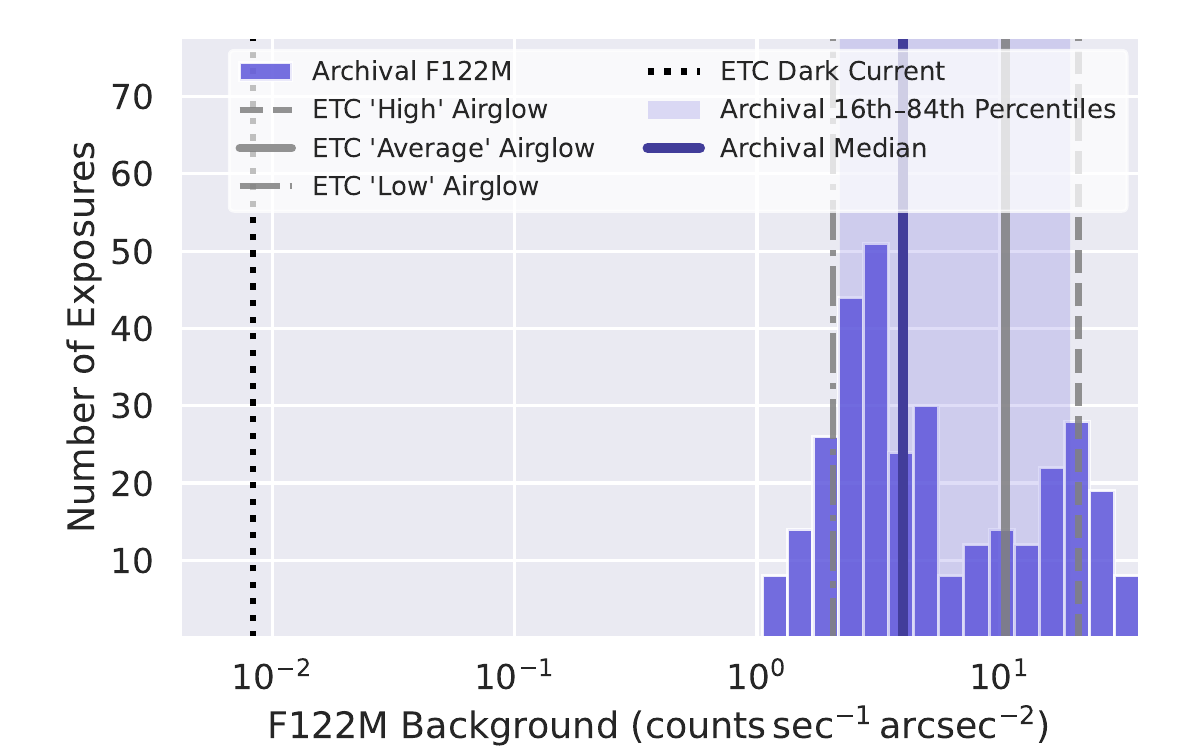}
\includegraphics[width=0.475\textwidth]{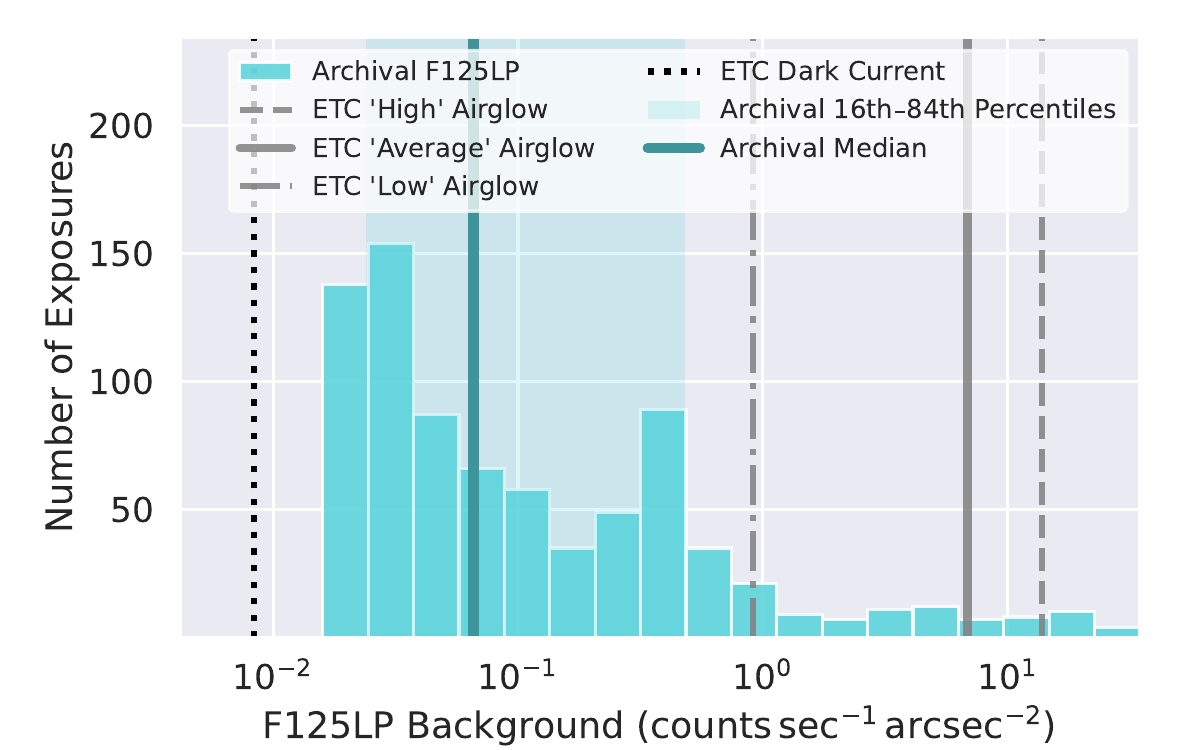}
\includegraphics[width=0.475\textwidth]{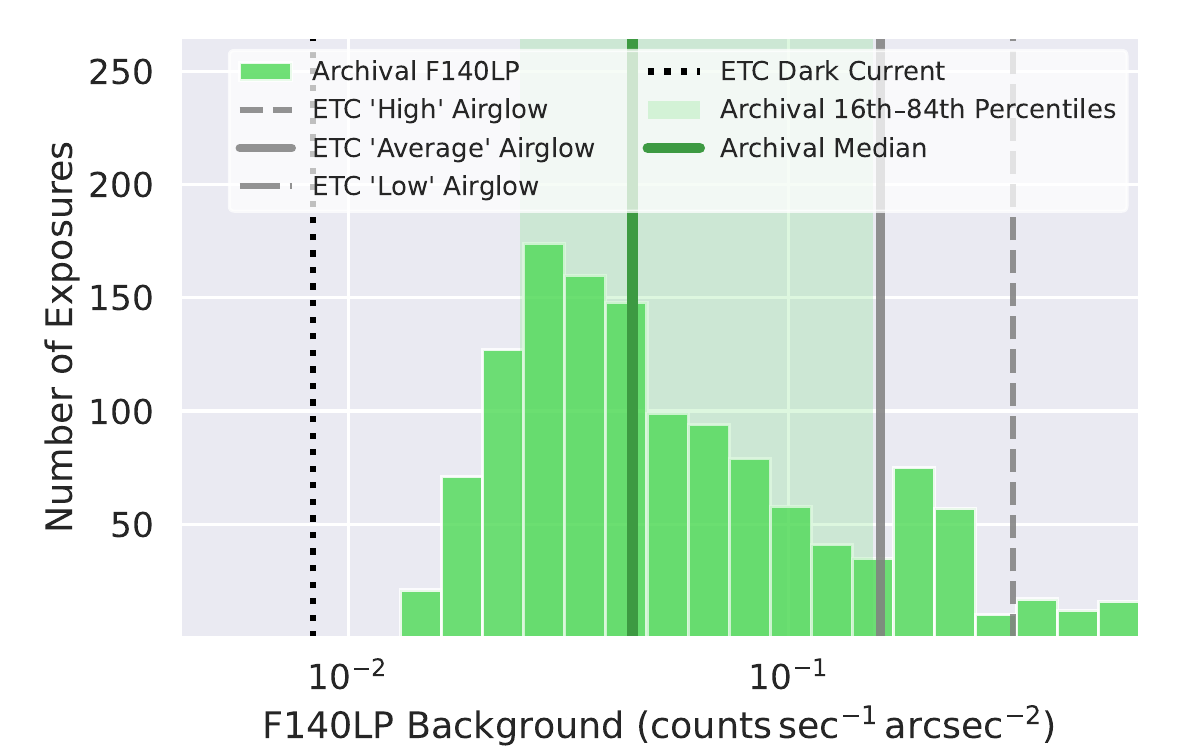}
\includegraphics[width=0.475\textwidth]{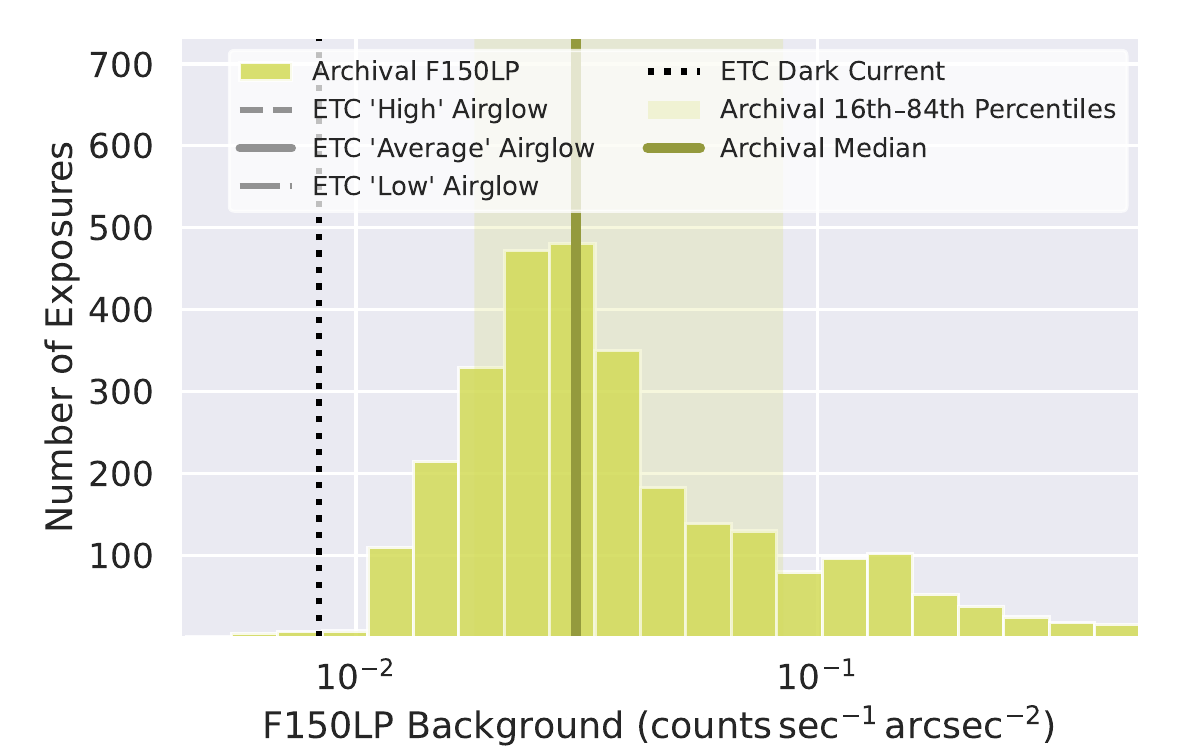}
\includegraphics[width=0.475\textwidth]{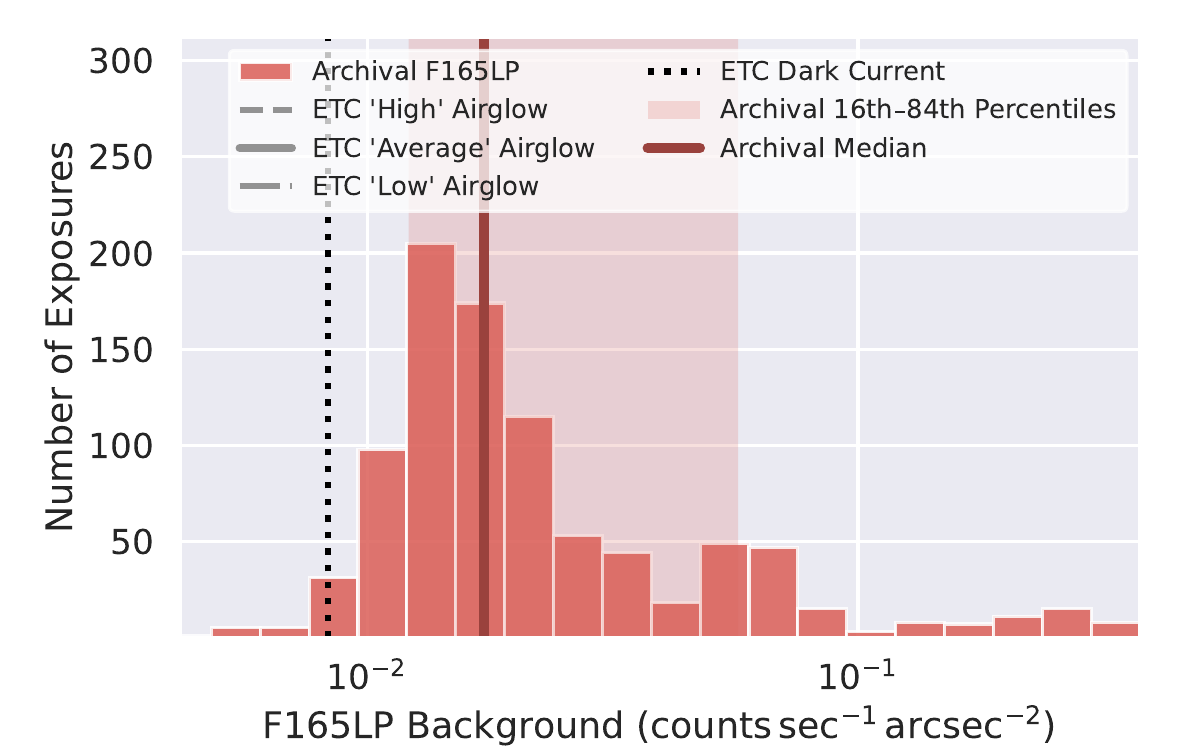}
\includegraphics[width=0.475\textwidth]{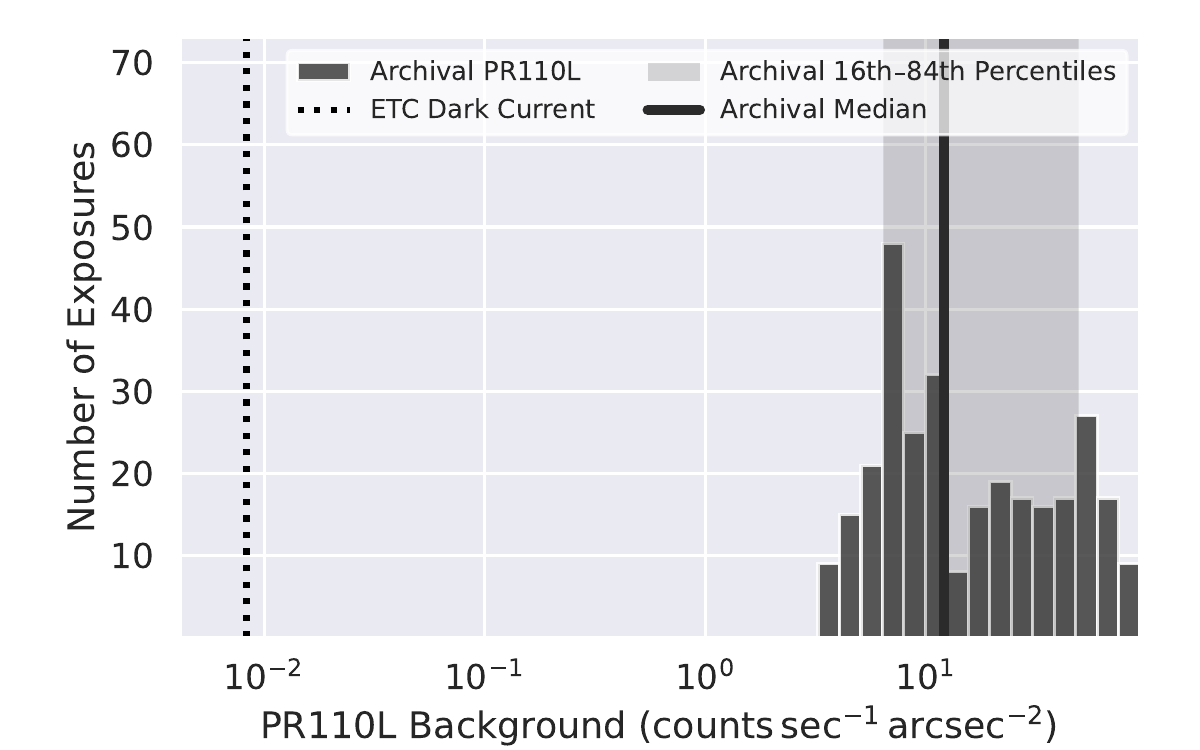}
\includegraphics[width=0.475\textwidth]{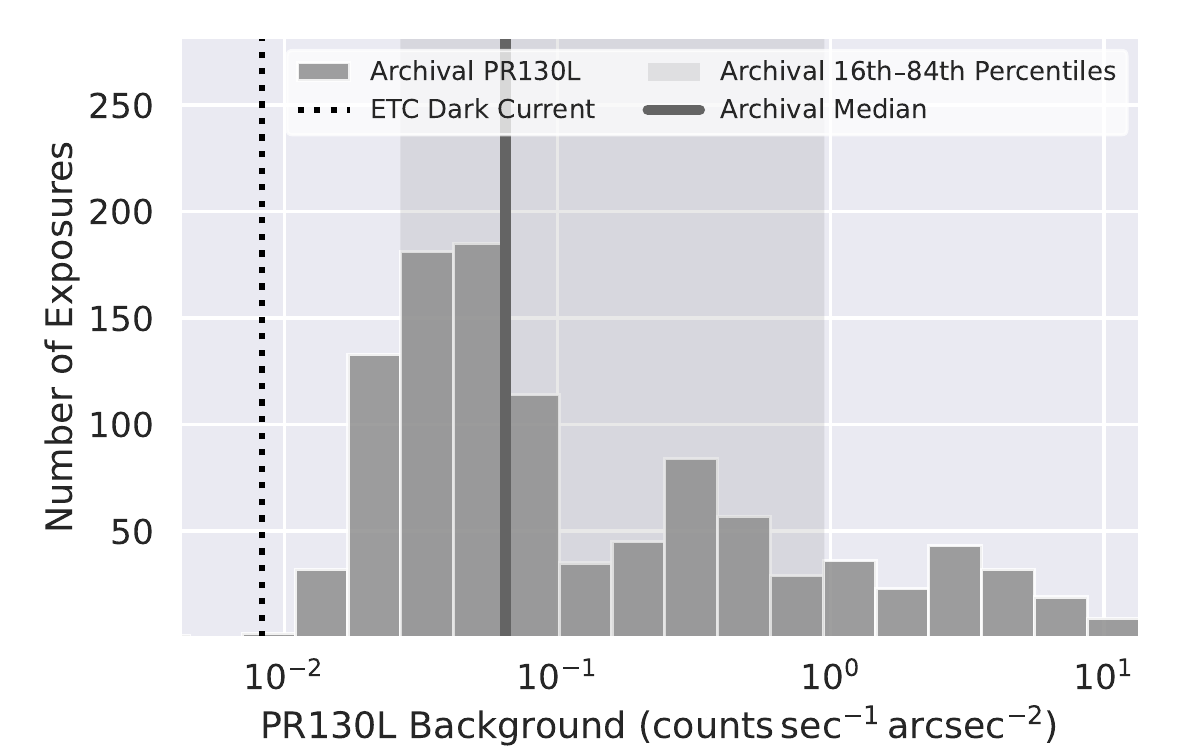}
\caption{Distributions of backgrounds measured in each filter. Median value is plotted for each, and 16\th\--84\th\ percentile ranges are shaded. Also shown for comparison are predicted ETC backgrounds for `low', `average', and `high' airglow (zodiacal light and Earthshine being negligible).  The standard assumed dark current rate (as of ETC v33.2) of 0.0082\,\cpspsqarc\ is also plotted, although this can vary over the detector. Prism measurements included for completeness; for instance, note that the background distribution for PR130L is very similar to that of F125LP, which has a very similar transmission function.}
\label{Fig:Background_Histograms}
\end{figure}

Having applied this dilation to each of our three masks generated in Sections~\ref{Subsection:Diffuse_Masking}, \ref{Subsection:Compact_Masking}, and \ref{Subsection:Tile_Masking}, we then combine these masks to produce our final mask. This combined mask is illustrated in panel H of Figures~\ref{Fig:Masking_Example_1}--\ref{Fig:Masking_Example_4}. We also show the dilated version of the tile-by-tile mask, in panel G of Figures~\ref{Fig:Masking_Example_1}--\ref{Fig:Masking_Example_4}, to show what the dilation looks like when applied to one of the three masks, before being combined into the final mask.

\subsection{Final Background Measurements}  \label{Subsection:Final_Background}

Having completed our masking process, the remaining un-masked portion of each frame should be a much better representation of the background. The un-masked images are shown in panel I of Figures~\ref{Fig:Masking_Example_1}--\ref{Fig:Masking_Example_4}. Even after the masking process, there can still be regions of elevated counts around the periphery of masked regions (for example, see panel I in Figure~\ref{Fig:Masking_Example_1}).  

For that reason, we once again consider our 5$\times$5 grid of 25 tiles, for computing our final measurement of the background. We throw out any tiles where \textgreater50\%\ of the pixels are masked, because those tiles will not necessarily be able to provide good statistics. Of the remaining tiles, we find the 5 tiles with the lowest mean count rates, as these tiles are presumably the least-contaminated with non-background counts. We combine the pixel values from these 5 tiles, and take the mean count rate as our final background level, in \cpspsqarc, for the exposure in question.

Percentile values for the distributions of backgrounds measured for all exposures in each filter are provided in Table~\ref{Table:Background_Percentiles}, along with predicted values from the ETC. Plots of the background distributions, again with predicted ETC backgrounds, are provided in Figure~\ref{Fig:Background_Histograms}. 

\section{Discrepancy Between Measured Backgrounds Values and ETC Predictions}  \label{Section:Discrepancy}

\begin{figure}
\centering
\includegraphics[width=0.475\textwidth]{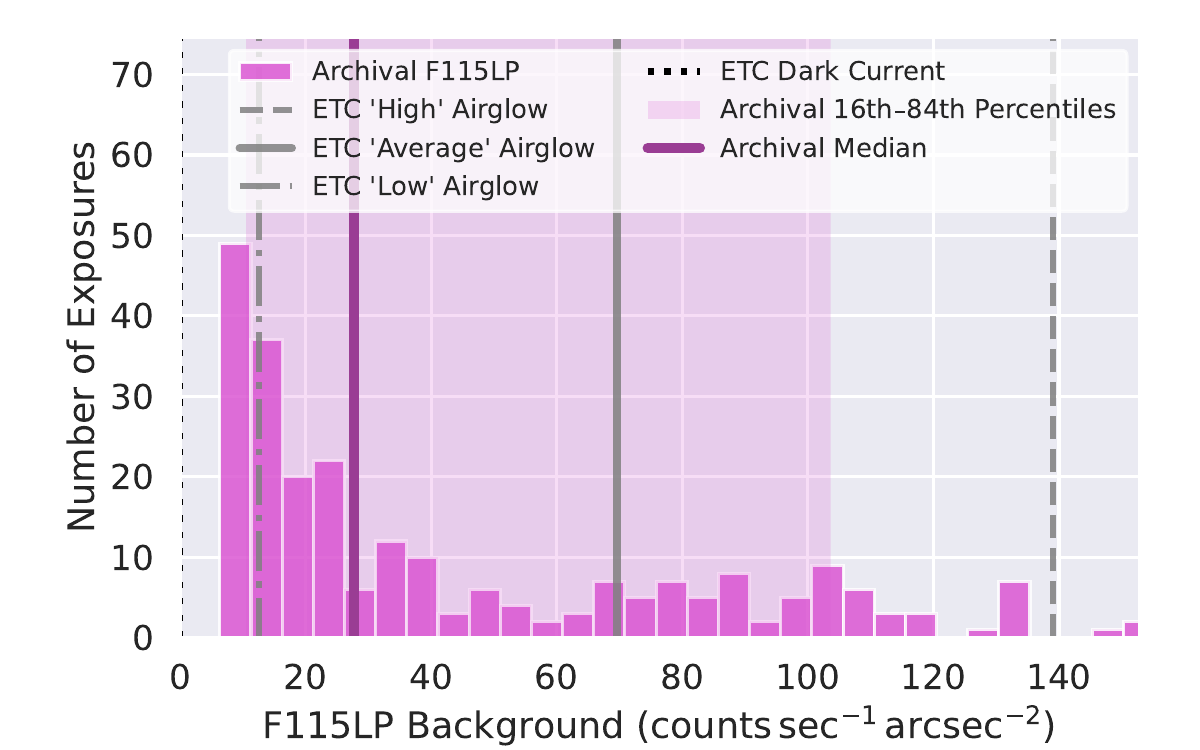}
\includegraphics[width=0.475\textwidth]{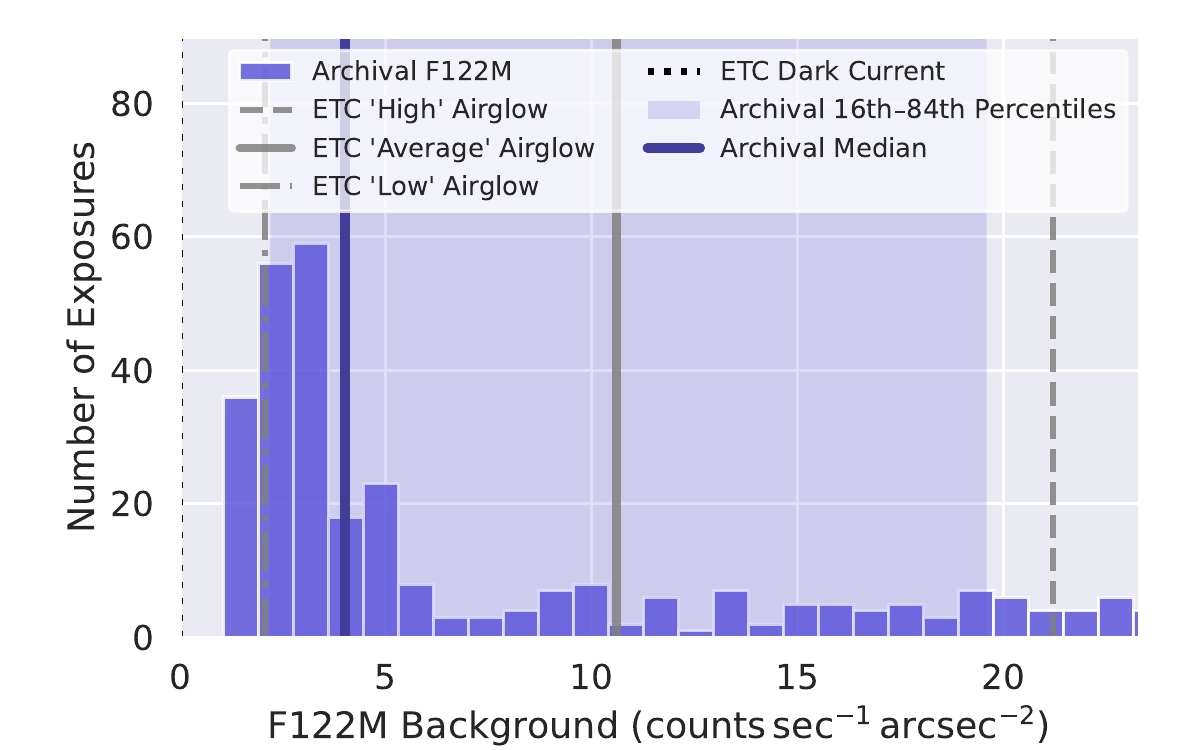}
\caption{Same as Figure~\ref{Fig:Background_Histograms} for F115LP and F122M, but plotted in linear space, to better illustrate the long tail towards higher background levels (which presents as a bimodality when plotted in logarithmic space).}
\label{Fig:Background_Histograms_Linear}
\end{figure}

Table~\ref{Table:Background_Percentiles} and Figure~\ref{Fig:Background_Histograms} make clear there there is a considerable discrepancy between the backgrounds predicted by the ETC, and what is actually measured in the archival data, for F115LP, F122M, F125LP, and F140LP. These are the filters where airglow is expected be the dominant source of the observed background. 

For F150LP and F165LP, the dark current rate is expected to dominate over the airglow. Indeed, we see that the standard assumed dark current rate (as of ETC v33.2) of 0.0082\,\cpspsqarc\ lies towards to lower end of the distributions of F150LP and F165LP backgrounds we measure in the data. If the dark rate were constant across the detector, we would expect 0.0082\,\cpspsqarc\ to be the floor level for measured backgrounds; however dark current rate varies over the detector (especially as a function of temperature), so in practice it can be below the 0.0082\,\cpspsqarc\ reference value. 

For the filters affected by airglow, however, the background levels we measure from the data are considerably lower than what the ETC predicts. The ETC predictions for `average' airglow are greater than the median of our measured background values by {\bf factors of 2.51 for F115LP, 2.64 for F122M, 105 for F125LP, and 3.64 for F140LP}. The implication of this is that the ETC is significantly over-estimating the airglow that is encountered in practice. 

We also note that the background value distributions of the airglow-dominated filters in Figure~\ref{Fig:Background_Histograms} appear bimodal, with a smaller peak at higher background levels -- although, given these histograms are plotted in logarithmic spacing, this second `peak' manifests as long tail in linear space, as shown in Figure~\ref{Fig:Background_Histograms_Linear}. We can find no obvious properties in common between observations that lie in this long tail. However, we note that the too-high backgrounds the ETC predicts for `average' airglow tend to lie in this long tail of high backgrounds.

\begin{table*}
\centering
\caption{Distribution of background levels measured for all eligible exposures for each SBC filter. Distributions provided as percentiles. Provided for comparison are backgrounds predicted by ETC version 33.2 the ETC, when assuming `low', `average', or `high' airglow. Zodiacal light and Earthshine both assumed to be average, as they have negligible impact on ETC predicted backgrounds at these wavelengths. All values in \cpspsqarc. The ETC background predictions do not incorporate the dark current rate, the standard assumed value of which (as of ETC version 33.2) is 0.008\,\cpspsqarc. The ETC does not provide imaging background levels for the prisms, so no such values are quoted here. }
\footnotesize
\label{Table:Background_Percentiles}
\begin{tabular}{lrrrrrrrr}
\toprule \toprule
\multicolumn{1}{c}{} &
\multicolumn{1}{c}{F115LP} &
\multicolumn{1}{c}{F122M} &
\multicolumn{1}{c}{F125LP} &
\multicolumn{1}{c}{F140LP} &
\multicolumn{1}{c}{F150LP} &
\multicolumn{1}{c}{F165LP} &
\multicolumn{1}{c}{PR110L} &
\multicolumn{1}{c}{PR130L} \\
\cmidrule(lr){2-9}
\makecell[l]{\bf Measured\\ \bf Percentiles} & &&&&&&& \\
5\th  & 7.65 & 1.64 & 0.0199 & 0.0189 & 0.0132 & 0.00948 & 4.48 & 0.0183 \\
10\th & 9.44 & 1.89 & 0.0217 & 0.0219 & 0.0155 & 0.0108  & 5.47 & 0.0226 \\
16\th & 10.5 & 2.20 & 0.0239 & 0.0246 & 0.0180 & 0.0121  & 6.41 & 0.0266 \\
50\th & 27.7 & 4.01 & 0.0658 & 0.0442 & 0.0300 & 0.0173  & 12.1 & 0.0646 \\
84\th & 104  & 19.6 & 0.480  & 0.155  & 0.0842 & 0.0569  & 49.3 & 0.948  \\
90\th & 129  & 22.8 & 0.916  & 0.209  & 0.132  & 0.0708  & 56.3 & 2.47   \\
95\th & 163  & 27.2 & 4.33   & 0.258  & 0.205  & 0.216   & 67.4 & 4.54   \\
\cmidrule(lr){2-9}
\makecell[l]{\bf ETC v33.2\\ \bf Percentiles} & &&&&&&& \\
`Low' Airglow & 12.6 & 2.06 & 0.091 & 0.002 & 0.0001 & 0.0001 & - & - \\
`Average' Airglow & 69.6 & 10.6 & 6.88 & 0.161 & 0.0002 & 0.0002 & - & - \\
`High' Airglow & 139 & 21.2 & 13.8 & 0.322 & 0.0004 & 0.003 & - & - \\
\bottomrule
\end{tabular}
\end{table*}

\subsection{Impact of Airglow on Measured versus ETC Backgrounds}  \label{Subsection:Airglow}

\begin{figure}
\centering
\includegraphics[width=0.975\textwidth]{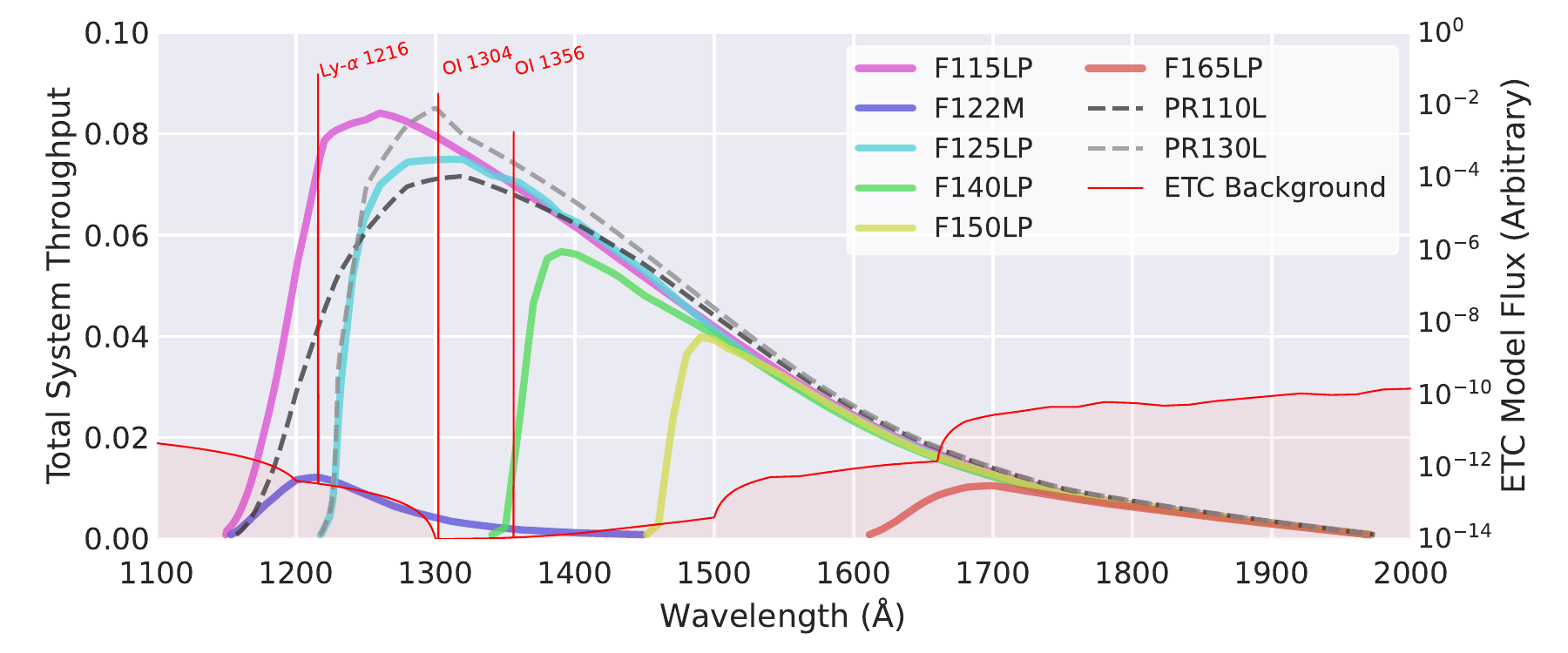}
\caption{Plot of total system throughput for all SBC filters, compared to the background model used by the ETC (v33.2), for `average' airglow, earthshine, and zodiacal light. Note that the throughput curves are plotted on a linear scale, whilst the ETC model is plotted on a logarithmic scale, for ease of viewing; note the $>$\,8 order of magnitude difference between line vs continuum brightness. The three airglow emission lines in the plotted ETC model are Ly-$\alpha$ at 1216\,\AA, and O{\sc i} at 1304\,\AA\ and 1356\,\AA.}
\label{Fig:Filters_and_Airglow}
\end{figure}

It seems possible that the ETC is over-estimating the contribution of airglow lines to the expected backgrounds. Figure~\ref{Fig:Filters_and_Airglow} shows the throughputs of the SBC filters, as compared to the wavelength-dependent model the ETC uses for the background (as of v33.2), consisting of airglow, earthshine, and zodiacal light\footnote{The ETC background model does not include the contribution of dark rate, which is handled separately in ETC calculations, as of v33.2; see Section~\ref{Subsection:Dark_vs_Background}.}. The continuum component of the ETC model is effectively negligible in this regime, with integrated flux \textless\,1\% that of the faintest included airglow line\footnote{Whilst F150LP and F165LP do not sample any of the significant airglow lines, the dark rate invariably dominates over earthshine and zodiacal light for these filters.}. 

The O{\sc i} 1356\,\AA\  airglow line, in particular, is sampled by all of the filters which have large discrepancies between the ETC and measured backgrounds. If the ETC issue is with the O{\sc i} lines, it could also explain why the background disagreement is so much stronger for F125LP. This is because F115LP and F122M sample Ly-$\alpha$ at 1216\,\AA, which  dominates the airglow, therefore reducing the relative impact of issue with O{\sc i}. Similarly F140LP is only slightly sensitive to O{\sc i} 1356\,\AA, as it lies in the filter's blue-end tail-off; this would also limit the relative impact of an issue with this line in the mode. F125LP, however, is fully sensitive to both main O{\sc i} lines, and excludes Ly-$\alpha$, making it highly affected by changes in the strengths of the O{\sc i} airglow. 

As an initial test of this hypothesis, we tried rescaling the airglow lines used the standard background model assumed by the ETC (v33.2), using `average' airglow, earthshine, and zodiacal light (as plotted in Figure~\ref{Fig:Filters_and_Airglow}). We set up an MCMC model to apply scaling factors to the strength of the three lines considered by the ETC in this wavelength range -- Ly-$\alpha$ 1216\,\AA, O{\sc i} 1304\,\AA, and O{\sc i} 1356\,\AA\ -- in order to find what combination of scaling factors would give reduced backgrounds that best fit our measured values\footnote{We also investigated rescaling the earthshine and zodiacal light components, however this had no impact on the best-fit rescaling factors for the airglow lines which entirely dominate the background in the ETC model.}. 

We found that applying rescaling factors of 0.41 to Ly-$\alpha$ 1216\,\AA, 0.0 to O{\sc i} 1304\,\AA, and 0.39 to O{\sc i} 1356\,\AA, gives backgrounds to agree with 5\%\ of our measured values for F115LP, F122M, and F140LP. However, even by applying the unlikely-seeming best-fit rescaling factor of 0.0 to the O{\sc i} 1304\,\AA, line (i.e., essentially removing the 1304\,\AA\ line), we are only able to reduce the F125LP background by a factor of 20 relative to the ETC prediction, not the factor of 105 we measure in practice\footnote{A factor of 20 reduction in the ETC prediction of the F125LP background corresponds to the 75\th\ percentile in the background distribution we measure in practice.}. It seems unlikely that the 1304 O{\sc i} line is so weak as to be effectively absent in most F125LP observation. But we note that there are a number of FUV airglow lines (such as the Lyman-Birge-Hopfield system of  N$_{2}$ lines) not currently included in the ETC model, which could account for this apparently-unphysical solution.

\section{Conclusions and Guidance for Users} \label{Section:Conclusion}

We have used 23 years of ACS/SBC data to study what background levels are encountered in practice, how much they vary, and what drives this variation.

We developed a method to automatically and consistently measure background levels in the 8,640 suitable archival SBC observations (Section~\ref{Section:Background_Measurement}). 

The backgrounds seen in SBC data vary considerably, with F115LP, F122M, F125LP, PR110L, and PR130L all showing at least an order of magnitude variation between the 10\th\ and 90\th\ percentile backgrounds. The F150LP and F165LP filters, however, show much less variation -- these are the only filters where the background is dominated by dark rate, not airglow. It appears that airglow variations are responsible for the highly variable backgrounds in the other filters.

\subsection{Background Differences vs ETC} \label{Subsection:Conc_ETC_Differences}

For the filters where the background is generally dominated by airglow, the background levels we measure from the data are significantly lower than what the ETC predicts (as of ETC v33.2). The ETC predictions for `average' airglow are greater than the median of our measured background values by factors of 2.51, 2.64, 105, and  3.64, for F115LP, F122M, F125LP, and F140LP, respectively. Our working hypothesis is that this difference is due to the O{\sc i} airglow lines at 1304\,\AA, and 1356\,\AA\ generally being fainter than expected by the ETC.

{\bf This indicates that the shorter-wavelength SBC filters can conduct observations with much more sensitivity in practice than had previously been expected}. As of ETC v34.1, a new option will be included for SBC calculations, where users can use the empirical background levels we report here, as opposed to the backgrounds predicted by the previous ETC model. These empirical backgrounds incorporate the contribution of the dark rate\footnote{Because the emprical ETC backgrounds incorporate the dark rate, the ETC no longer adds a dark rate component in addition to the background (as it usually would) when using the empirical background option, to prevent double-counting.}; however, the filters where the dark rate is most consequential (F150LP and F165LP) are also least affected by changes in expected airglow, meaning users should not need to worry about disentangling their respective contributions.

\subsection{Bright Object Limits} \label{Subsection:Conc_Bright_Object_Limits}

Our re-evaluation of expected backgrounds has implications when considering bright object protection. Because MAMA detectors like the SBC can be catastrophically and permanently damaged by excessive count rates, the SBC imposes bright object protection limits (see \citealp{Ryon2019D}, and the ACS Instrument Handbook; \citealp{Hubble-ACS2024}, Section 7.2\footnote{\url{https://hst-docs.stsci.edu/acsihb/chapter-7-observing-techniques/7-2-sbc-bright-object-protection}}). If the SBC electronics detect an excessive count rate, then an observation is immediately halted (by closing the shutter, or cutting voltage to the detector).

The SBC hard screening limits are 50\,counts sec$^{-1}$ pixel$^{-1}$ for any given pixel, and 200,000\,counts sec$^{-1}$ globally. Integrated over the 1,055\,arcsec$^{2}$ SBC field of view, the global screening limit corresponds to 189\,\cpspsqarc\  (given the standard 0.030$^{\prime\prime}$$\times$0.034$^{\prime\prime}$ pixel area). 

In F115LP, the ETC airglow model (as of ETC v33.2) predicts backgrounds that reach 37\%\ of the global screening rate for `average' airglow, and 74\%\ of the global screening rate for `high' airglow. Similarly, for F122M, the `high' airglow ETC predicted background corresponds to 11\%\ of the global screening rate.  These are large enough fractions of the screening limit that they could affect users' observing strategy decisions.

However, the empirical background level measurements we report here mean that a median background in F115LP would only reach 15\%\ of the global screening limit (and only 2\%\ for F122M). Whilst a 90\th\ percentile F115LP background would reach 68\%\ of the limit, we also note that for all of the archival SBC exposures considered in this work, the background {\it never} exceeded 50\,\cpspsqarc\ -- 26\%\ of the global screening limit -- when the Sun was beneath the horizon (i.e., corresponding to the \texttt{SHADOW} special requirement). 

Users should can this information when accounting for whether the background is likely to contribute to their observation's risk of approaching the screening rate.

\subsection{Predicting Backgrounds} \label{Subsection:Conc_Predicting}

In ACS ISR 2026-01 (Clark et al., 2025), the companion report to this one, we use a wide range of parameters associated with each observation (telescope ephemeris, geocoronal conditions, instrument temperature, etc) to {\it predict} the background expected for a given observation. Users interested in considering how observations can be planned so as to minimize background levels are directed to read that ISR.

\section{Acknowledgements}

This research made use of \texttt{Astropy}\footnote{\url{https://www.astropy.org/}}, a community-developed core \texttt{Python} package for Astronomy \citep{astropy2013,astropy2019}. This research made use of Photutils\footnote{\url{https://photutils.readthedocs.io}}, an \texttt{Astropy}-affiliated package for
detection and photometry of astronomical sources\citep{Bradley2020C}. This research made use of \texttt{NumPy}\footnote{\url{https://numpy.org/}} \citep{VanDerWalt2011B,Harris2020A}, \texttt{SciPy}\footnote{\url{https://scipy.org/}} \citep{SciPy2001,SciPy2020}, and \texttt{Matplotlib}\footnote{\url{https://matplotlib.org/}} \citep{Hunter2007A}. This research made use of the \texttt{pandas}\footnote{\url{https://pandas.pydata.org/}} data structures package for \texttt{Python} \citep{McKinney2010}. This research made use of \texttt{corner}\footnote{\url{https://corner.readthedocs.io}}, a python package for the display of multidimensional samples \citep{ForemanMackey2016D}. This research made use of \texttt{iPython}, an enhanced interactive \texttt{Python} \citep{Perez2007A}. This research made use of \texttt{TOPCAT}\footnote{\url{http://www.star.bris.ac.uk/~mbt/topcat/}} \citep{Taylor2005A}, an interactive graphical viewer and editor for tabular data.

\vspace{3.5ex plus 1ex minus 0.2ex}
\appendix
\newpage 

{\normalsize 
\setlength{\bibsep}{1.0pt} 
\def\ref@jnl#1{{\rmfamily #1}}%
\newcommand\aj{\ref@jnl{AJ}}%
\newcommand\araa{\ref@jnl{ARA\&A}}%
\newcommand\apj{\ref@jnl{ApJ}}%
\newcommand\apjl{\ref@jnl{ApJ}}%
\newcommand\apjs{\ref@jnl{ApJS}}%
\newcommand\ao{\ref@jnl{Appl.~Opt.}}%
\newcommand\apss{\ref@jnl{Ap\&SS}}%
\newcommand\aap{\ref@jnl{A\&A}}%
\newcommand\aapr{\ref@jnl{A\&A~Rev.}}%
\newcommand\aaps{\ref@jnl{A\&AS}}%
\newcommand\azh{\ref@jnl{AZh}}%
\newcommand\baas{\ref@jnl{BAAS}}%
\newcommand\jrasc{\ref@jnl{JRASC}}%
\newcommand\memras{\ref@jnl{MmRAS}}%
\newcommand\mnras{\ref@jnl{MNRAS}}%
\newcommand\pra{\ref@jnl{Phys.~Rev.~A}}%
\newcommand\prb{\ref@jnl{Phys.~Rev.~B}}%
\newcommand\prc{\ref@jnl{Phys.~Rev.~C}}%
\newcommand\prd{\ref@jnl{Phys.~Rev.~D}}%
\newcommand\pre{\ref@jnl{Phys.~Rev.~E}}%
\newcommand\prl{\ref@jnl{Phys.~Rev.~Lett.}}%
\newcommand\pasp{\ref@jnl{PASP}}%
\newcommand\pasj{\ref@jnl{PASJ}}%
\newcommand\qjras{\ref@jnl{QJRAS}}%
\newcommand\skytel{\ref@jnl{S\&T}}%
\newcommand\solphys{\ref@jnl{Sol.~Phys.}}%
\newcommand\sovast{\ref@jnl{Soviet~Ast.}}%
\newcommand\ssr{\ref@jnl{Space~Sci.~Rev.}}%
\newcommand\zap{\ref@jnl{ZAp}}%
\newcommand\nat{\ref@jnl{Nature}}%
\newcommand\iaucirc{\ref@jnl{IAU~Circ.}}%
\newcommand\aplett{\ref@jnl{Astrophys.~Lett.}}%
\newcommand\apspr{\ref@jnl{Astrophys.~Space~Phys.~Res.}}%
\newcommand\bain{\ref@jnl{Bull.~Astron.~Inst.~Netherlands}}%
\newcommand\fcp{\ref@jnl{Fund.~Cosmic~Phys.}}%
\newcommand\gca{\ref@jnl{Geochim.~Cosmochim.~Acta}}%
\newcommand\grl{\ref@jnl{Geophys.~Res.~Lett.}}%
\newcommand\jcp{\ref@jnl{J.~Chem.~Phys.}}%
\newcommand\jgr{\ref@jnl{J.~Geophys.~Res.}}%
\newcommand\jqsrt{\ref@jnl{J.~Quant.~Spec.~Radiat.~Transf.}}%
\newcommand\memsai{\ref@jnl{Mem.~Soc.~Astron.~Italiana}}%
\newcommand\nphysa{\ref@jnl{Nucl.~Phys.~A}}%
\newcommand\physrep{\ref@jnl{Phys.~Rep.}}%
\newcommand\physscr{\ref@jnl{Phys.~Scr}}%
\newcommand\planss{\ref@jnl{Planet.~Space~Sci.}}%
\newcommand\procspie{\ref@jnl{Proc.~SPIE}}%
\newcommand\pasa{\ref@jnl{PASA}}%

\bibliographystyle{mnras}
\bibliography{ChrisBib}}

@ARTICLE{Bertin1996A,
   author = {{Bertin}, E. and {Arnouts}, S.},
    title = "{SExtractor: Software for source extraction.}",
  journal = {\aaps},
     year = 1996,
    month = jun,
   volume = 117,
    pages = {393-404},
   adsurl = {http://adsabs.harvard.edu/abs/1996A%26AS..117..393B}
}

@ARTICLE{Astropy2013,
   author = {{Astropy Collaboration} and {Robitaille}, T.~P. and {Tollerud}, E.~J. and
	{Greenfield}, P. and {Droettboom}, M. and {Bray}, E. and {Aldcroft}, T. and
	{Davis}, M. and {Ginsburg}, A. and {Price-Whelan}, A.~M. and
	{Kerzendorf}, W.~E. and {Conley}, A. and {Crighton}, N. and
	{Barbary}, K. and {Muna}, D. and {Ferguson}, H. and {Grollier}, F. and
	{Parikh}, M.~M. and {Nair}, P.~H. and {Unther}, H.~M. and {Deil}, C. and
	{Woillez}, J. and {Conseil}, S. and {Kramer}, R. and {Turner}, J.~E.~H. and
	{Singer}, L. and {Fox}, R. and {Weaver}, B.~A. and {Zabalza}, V. and
	{Edwards}, Z.~I. and {Azalee Bostroem}, K. and {Burke}, D.~J. and
	{Casey}, A.~R. and {Crawford}, S.~M. and {Dencheva}, N. and
	{Ely}, J. and {Jenness}, T. and {Labrie}, K. and {Lim}, P.~L. and
	{Pierfederici}, F. and {Pontzen}, A. and {Ptak}, A. and {Refsdal}, B. and
	{Servillat}, M. and {Streicher}, O.},
    title = "{Astropy: A community Python package for astronomy}",
  journal = {\aap},
archivePrefix = "arXiv",
   eprint = {1307.6212},
 primaryClass = "astro-ph.IM",
     year = 2013,
    month = oct,
   volume = 558,
      eid = {A33},
    pages = {A33},
      doi = {10.1051/0004-6361/201322068},
   adsurl = {http://adsabs.harvard.edu/abs/2013A%26A...558A..33A}
}

@INPROCEEDINGS{Taylor2005A,
   author = {{Taylor}, M.~B.},
    title = "{TOPCAT {\amp} STIL: Starlink Table/VOTable Processing Software}",
booktitle = {Astronomical Data Analysis Software and Systems XIV},
     year = 2005,
   series = {Astronomical Society of the Pacific Conference Series},
   volume = 347,
   editor = {{Shopbell}, P. and {Britton}, M. and {Ebert}, R.},
    month = dec,
    pages = {29},
   adsurl = {http://adsabs.harvard.edu/abs/2005ASPC..347...29T}
}

@ARTICLE{VanDerWalt2011B,
   author  =  {Stéfan van der Walt and S. Chris Colbert and Gaël Varoquaux},
    title  =  {The NumPy Array: A Structure for Efficient Numerical Computation},
  journal  =  {Computing in Science \& Engineering},
   volume  =  {13},
   number  =  {2},
    pages  =  {22-30},
     year  =  {2011},
      doi  =  {10.1109/MCSE.2011.37},
      URL  =  {http://aip.scitation.org/doi/abs/10.1109/MCSE.2011.37},
   eprint  =  {http://aip.scitation.org/doi/pdf/10.1109/MCSE.2011.37}
}

@ARTICLE{Hunter2007A,
   author  =  {John D. Hunter},
    title  =  {Matplotlib: A 2D Graphics Environment},
  journal  =  {Computing in Science \& Engineering},
   volume  =  {9},
   number  =  {3},
    pages  =  {90-95},
     year  =  {2007},
      doi  =  {10.1109/MCSE.2007.55},
      URL  =  {http://aip.scitation.org/doi/abs/10.1109/MCSE.2007.55},
eprint  =  {http://aip.scitation.org/doi/pdf/10.1109/MCSE.2007.55}
}

@ARTICLE{ForemanMackey2016D,
       author = {{Foreman-Mackey}, Daniel},
        title = "{corner.py: Scatterplot matrices in Python}",
      journal = {The Journal of Open Source Software},
         year = 2016,
        month = Jun,
       volume = {1},
        pages = {24},
          doi = {10.21105/joss.00024},
       adsurl = {https://ui.adsabs.harvard.edu/#abs/2016JOSS....1...24F}
}

@InProceedings{McKinney2010,
  author      = { Wes McKinney },
  title       = { Data Structures for Statistical Computing in Python },
  booktitle   = { Proceedings of the 9th Python in Science Conference },
  pages       = { 51 - 56 },
  year        = { 2010 },
  editor      = { St\'efan van der Walt and Jarrod Millman }
}

@Misc{SciPy2001,
      author = {Eric Jones and Travis Oliphant and Pearu Peterson and others},
       title = {{SciPy}: Open source scientific tools for {Python}},
        year = {2001},
         url = "http://www.scipy.org/"
}

@Article{Perez2007A,
    Author    = {P\'erez, Fernando and Granger, Brian E.},
    Title     = {{IP}ython: a System for Interactive Scientific Computing},
    Journal   = {Computing in Science and Engineering},
    Volume    = {9},
    Number    = {3},
    Pages     = {21--29},
    month     = may,
    year      = 2007,
    url       = "https://ipython.org",
    ISSN      = "1521-9615",
    doi       = {10.1109/MCSE.2007.53},
    publisher = {IEEE Computer Society},
}

@ARTICLE{Astropy2019,
       author = {{Astropy Collaboration} and {Price-Whelan}, A.~M. and
         {Sip{\H{o}}cz}, B.~M. and {G{\"u}nther}, H.~M. and {Lim}, P.~L. and
         {Crawford}, S.~M. and {Conseil}, S. and {Shupe}, D.~L. and
         {Craig}, M.~W. and {Dencheva}, N. and {Ginsburg}, A. and {Vand
        erPlas}, J.~T. and {Bradley}, L.~D. and {P{\'e}rez-Su{\'a}rez}, D. and
         {de Val-Borro}, M. and {Aldcroft}, T.~L. and {Cruz}, K.~L. and
         {Robitaille}, T.~P. and {Tollerud}, E.~J. and {Ardelean}, C. and
         {Babej}, T. and {Bach}, Y.~P. and {Bachetti}, M. and {Bakanov}, A.~V. and
         {Bamford}, S.~P. and {Barentsen}, G. and {Barmby}, P. and
         {Baumbach}, A. and {Berry}, K.~L. and {Biscani}, F. and {Boquien}, M. and
         {Bostroem}, K.~A. and {Bouma}, L.~G. and {Brammer}, G.~B. and
         {Bray}, E.~M. and {Breytenbach}, H. and {Buddelmeijer}, H. and
         {Burke}, D.~J. and {Calderone}, G. and {Cano Rodr{\'\i}guez}, J.~L. and
         {Cara}, M. and {Cardoso}, J.~V.~M. and {Cheedella}, S. and {Copin}, Y. and
         {Corrales}, L. and {Crichton}, D. and {D'Avella}, D. and {Deil}, C. and
         {Depagne}, {\'E}. and {Dietrich}, J.~P. and {Donath}, A. and
         {Droettboom}, M. and {Earl}, N. and {Erben}, T. and {Fabbro}, S. and
         {Ferreira}, L.~A. and {Finethy}, T. and {Fox}, R.~T. and
         {Garrison}, L.~H. and {Gibbons}, S.~L.~J. and {Goldstein}, D.~A. and
         {Gommers}, R. and {Greco}, J.~P. and {Greenfield}, P. and
         {Groener}, A.~M. and {Grollier}, F. and {Hagen}, A. and {Hirst}, P. and
         {Homeier}, D. and {Horton}, A.~J. and {Hosseinzadeh}, G. and {Hu}, L. and
         {Hunkeler}, J.~S. and {Ivezi{\'c}}, {\v{Z}}. and {Jain}, A. and
         {Jenness}, T. and {Kanarek}, G. and {Kendrew}, S. and {Kern}, N.~S. and
         {Kerzendorf}, W.~E. and {Khvalko}, A. and {King}, J. and {Kirkby}, D. and
         {Kulkarni}, A.~M. and {Kumar}, A. and {Lee}, A. and {Lenz}, D. and
         {Littlefair}, S.~P. and {Ma}, Z. and {Macleod}, D.~M. and
         {Mastropietro}, M. and {McCully}, C. and {Montagnac}, S. and
         {Morris}, B.~M. and {Mueller}, M. and {Mumford}, S.~J. and {Muna}, D. and
         {Murphy}, N.~A. and {Nelson}, S. and {Nguyen}, G.~H. and
         {Ninan}, J.~P. and {N{\"o}the}, M. and {Ogaz}, S. and {Oh}, S. and
         {Parejko}, J.~K. and {Parley}, N. and {Pascual}, S. and {Patil}, R. and
         {Patil}, A.~A. and {Plunkett}, A.~L. and {Prochaska}, J.~X. and
         {Rastogi}, T. and {Reddy Janga}, V. and {Sabater}, J. and
         {Sakurikar}, P. and {Seifert}, M. and {Sherbert}, L.~E. and
         {Sherwood-Taylor}, H. and {Shih}, A.~Y. and {Sick}, J. and
         {Silbiger}, M.~T. and {Singanamalla}, S. and {Singer}, L.~P. and
         {Sladen}, P.~H. and {Sooley}, K.~A. and {Sornarajah}, S. and
         {Streicher}, O. and {Teuben}, P. and {Thomas}, S.~W. and
         {Tremblay}, G.~R. and {Turner}, J.~E.~H. and {Terr{\'o}n}, V. and
         {van Kerkwijk}, M.~H. and {de la Vega}, A. and {Watkins}, L.~L. and
         {Weaver}, B.~A. and {Whitmore}, J.~B. and {Woillez}, J. and
         {Zabalza}, V. and {Astropy Contributors}},
        title = "{The Astropy Project: Building an Open-science Project and Status of the v2.0 Core Package}",
      journal = {\aj},
         year = "2018",
        month = "Sep",
       volume = {156},
       number = {3},
          eid = {123},
        pages = {123},
          doi = {10.3847/1538-3881/aabc4f},
archivePrefix = {arXiv},
       eprint = {1801.02634},
 primaryClass = {astro-ph.IM},
       adsurl = {https://ui.adsabs.harvard.edu/abs/2018AJ....156..123A}
}

@ARTICLE{Harris2020A,
       author = {{Harris}, Charles R. and {Jarrod Millman}, K. and
         {van der Walt}, St{\'e}fan J. and {Gommers}, Ralf and
         {Virtanen}, Pauli and {Cournapeau}, David and {Wieser}, Eric and
         {Taylor}, Julian and {Berg}, Sebastian and {Smith}, Nathaniel J. and
         {Kern}, Robert and {Picus}, Matti and {Hoyer}, Stephan and
         {van Kerkwijk}, Marten H. and {Brett}, Matthew and {Haldane}, Allan and
         {Fern{\'a}ndez del R{\'\i}o}, Jaime and {Wiebe}, Mark and
         {Peterson}, Pearu and {G{\'e}rard-Marchant}, Pierre and
         {Sheppard}, Kevin and {Reddy}, Tyler and {Weckesser}, Warren and
         {Abbasi}, Hameer and {Gohlke}, Christoph and {Oliphant}, Travis E.},
        title = "{Array Programming with NumPy}",
      journal = {\nat},
         year = 2020,
        month = jun,
       volume = {585},
       number = {7825},
        pages = {537-362},
          doi = {https://doi.org/10.1038/s41586-020-2649-2},
archivePrefix = {arXiv},
       eprint = {2006.10256},
 primaryClass = {cs.MS},
       adsurl = {https://ui.adsabs.harvard.edu/abs/2020arXiv200610256H}
}

@MISC{Bradley2020C,
       author = {{Bradley}, Larry and {Sip{\H{o}}cz}, Brigitta and {Robitaille}, Thomas and {Tollerud}, Erik and {Vin{\'\i}cius}, Z{\'e} and {Deil}, Christoph and {Barbary}, Kyle and {Wilson}, Tom J and {Busko}, Ivo and {G{\"u}nther}, Hans Moritz and {Cara}, Mihai and {Conseil}, Simon and {Bostroem}, Azalee and {Droettboom}, Michael and {Bray}, E.~M. and {Andersen Bratholm}, Lars and {Lim}, P.~L. and {Barentsen}, Geert and {Craig}, Matt and {Pascual}, Sergio and {Perren}, Gabriel and {Greco}, Johnny and {Donath}, Axel and {De Val-Borro}, Miguel and {Kerzendorf}, Wolfgang and {Bach}, Yoonsoo P. and {Weaver}, Benjamin Alan and {D'Eugenio}, Francesco and {Souchereau}, Harrison and {Ferreira}, Leonardo},
        title = "{astropy/photutils: 1.0.1}",
         year = 2020,
        month = sep,
          eid = {10.5281/zenodo.596036},
          doi = {10.5281/zenodo.596036},
      version = {1.0.1},
    publisher = {Zenodo},
       adsurl = {https://ui.adsabs.harvard.edu/abs/2020zndo....596036B}
}

@ARTICLE{SciPy2020,
  author  = {Virtanen, Pauli and Gommers, Ralf and Oliphant, Travis E. and
            Haberland, Matt and Reddy, Tyler and Cournapeau, David and
            Burovski, Evgeni and Peterson, Pearu and Weckesser, Warren and
            Bright, Jonathan and {van der Walt}, St{\'e}fan J. and
            Brett, Matthew and Wilson, Joshua and Millman, K. Jarrod and
            Mayorov, Nikolay and Nelson, Andrew R. J. and Jones, Eric and
            Kern, Robert and Larson, Eric and Carey, C J and
            Polat, {\.I}lhan and Feng, Yu and Moore, Eric W. and
            {VanderPlas}, Jake and Laxalde, Denis and Perktold, Josef and
            Cimrman, Robert and Henriksen, Ian and Quintero, E. A. and
            Harris, Charles R. and Archibald, Anne M. and
            Ribeiro, Ant{\^o}nio H. and Pedregosa, Fabian and
            {van Mulbregt}, Paul and {SciPy 1.0 Contributors}},
  title   = {{{SciPy} 1.0: Fundamental Algorithms for Scientific
            Computing in Python}},
  journal = {Nature Methods},
  year    = {2020},
  volume  = {17},
  pages   = {261--272},
  adsurl  = {https://rdcu.be/b08Wh},
  doi     = {10.1038/s41592-019-0686-2},
}

@INPROCEEDINGS{Tran2003B,
       author = {{Tran}, H.~D. and {Meurer}, G.~R. and {Ford}, H.~C. and {Martel}, A.~R. and {Sirianni}, M. and {Bohlin}, R.~C. and {Clampin}, M. and {Cox}, C. and {de Marchi}, G. and {Hartig}, G.~F. and {Kimble}, R.~A. and {Argabright}, V.},
        title = "{On-Orbit Performance of the ACS Solar Blind Channel}",
    booktitle = {HST Calibration Workshop : Hubble after the Installation of the ACS and the NICMOS Cooling System},
         year = 2003,
       editor = {{Arribas}, Santiago and {Koekemoer}, Anton and {Whitmore}, Brad},
        month = jan,
        pages = {86},
       adsurl = {https://ui.adsabs.harvard.edu/abs/2003hstc.conf...86T}
}

@INPROCEEDINGS{Clampin2000B,
       author = {{Clampin}, Mark and {Ford}, Holland C. and {Bartko}, Frank and {Bely}, Pierre Y. and {Broadhurst}, Tom and {Burrows}, Christopher J. and {Cheng}, Edward S. and {Crocker}, James H. and {Franx}, Marijn and {Feldman}, Paul D. and {Golimowski}, David A. and {Hartig}, George F. and {Illingworth}, Garth and {Kimble}, Randy A. and {Lesser}, Michael P. and {Miley}, George H. and {Postman}, Marc and {Rafal}, Marc D. and {Rosati}, Piero and {Sparks}, William B. and {Tsvetanov}, Zlatan and {White}, Richard L. and {Sullivan}, Pamela and {Volmer}, Paul and {LaJeunesse}, Tom},
        title = "{Advanced camera for surveys}",
    booktitle = {UV, Optical, and IR Space Telescopes and Instruments},
         year = 2000,
       editor = {{Breckinridge}, James B. and {Jakobsen}, Peter},
       series = {Society of Photo-Optical Instrumentation Engineers (SPIE) Conference Series},
       volume = {4013},
        month = jul,
        pages = {344-351},
          doi = {10.1117/12.394016},
       adsurl = {https://ui.adsabs.harvard.edu/abs/2000SPIE.4013..344C}
}

@ARTICLE{Bahcall1982C,
       author = {{Bahcall}, J.~N. and {Spitzer}, Jr., L.},
        title = "{The Space Telescope}",
      journal = {Scientific American},
         year = 1982,
        month = jul,
       volume = {247},
        pages = {40-51},
          doi = {10.1038/scientificamerican0782-40},
       adsurl = {https://ui.adsabs.harvard.edu/abs/1982SciAm.247a..40B}
}

@MANUAL{Hubble-ACS2024,
       author = {{Stark}, D. V. and others},
        title = "{ACS Instrument Handbook}",
         year = "2024",
      edition = "24",
 organization = "Space Telescope Science Institute"
}

@MISC{Johnson2024C,
       author = {{Johnson}, Christian I. and {Lockwood}, Sean and {Gomez}, Sierra and {French}, David},
        title = "{Cycle 30 COS NUV Dark Monitor Summary}",
 howpublished = {Instrument Science Report COS 2024-14},
         year = 2024,
        month = sep,
        pages = {14},
       adsurl = {https://ui.adsabs.harvard.edu/abs/2024cos..rept...14J}
}

@ARTICLE{Putis2018A,
       author = {{Putis}, M. and {Bobik}, P. and {Mackovjak}, S.},
        title = "{Method for Analysis of the Effect of Geomagnetic Disturbances on Ultraviolet Airglow Intensity}",
      journal = {Earth and Space Science},
         year = 2018,
        month = nov,
       volume = {5},
       number = {11},
        pages = {790-800},
          doi = {10.1029/2017EA000358},
       adsurl = {https://ui.adsabs.harvard.edu/abs/2018E&SS....5..790P}
}

@ARTICLE{Waldrop2013A,
       author = {{Waldrop}, L. and {Paxton}, L.~J.},
        title = "{Lyman {\ensuremath{\alpha}} airglow emission: Implications for atomic hydrogen geocorona variability with solar cycle}",
      journal = {Journal of Geophysical Research (Space Physics)},
         year = 2013,
        month = sep,
       volume = {118},
       number = {9},
        pages = {5874-5890},
          doi = {10.1002/jgra.50496},
       adsurl = {https://ui.adsabs.harvard.edu/abs/2013JGRA..118.5874W}
}

@MISC{Avila2017B,
       author = {{Avila}, R.~J.},
        title = "{Updated Measurements of ACS/SBC Dark Rates}",
 howpublished = {Instrument Science Report ACS 2017-4, 8 pages},
         year = 2017,
        month = may,
        pages = {4},
       adsurl = {https://ui.adsabs.harvard.edu/abs/2017acs..rept....4A}
}

@ARTICLE{He2020B,
       author = {{He}, Fei and {Wei}, Yong and {Wan}, Weixing},
        title = "{Equatorial aurora: the aurora-like airglow in the negative magnetic anomaly}",
      journal = {National Science Review},
         year = 2020,
        month = oct,
       volume = {7},
       number = {10},
        pages = {1606-1615},
          doi = {10.1093/nsr/nwaa083},
       adsurl = {https://ui.adsabs.harvard.edu/abs/2020NSRev...7.1606H}
}

@MISC{Avila2018A,
       author = {{Avila}, Roberto J. and {Arslanian}, S. and {Bourque}, Matthew and {Eck}, Weston},
        title = "{Mitigating Elevated Dark Rates in SBC Imaging}",
 howpublished = {Instrument Science Report ACS 2018-07, 6 pages},
         year = 2018,
        month = oct,
        pages = {7},
       adsurl = {https://ui.adsabs.harvard.edu/abs/2018acs..rept....7A}
}

@software{Bradley2025B,
       author = {{Bradley}, Larry and {Sip{\H{o}}cz}, Brigitta and {Robitaille}, Thomas and {Tollerud}, Erik and {Vin{\'\i}cius}, Z{\'e} and {Deil}, Christoph and {Barbary}, Kyle and {Wilson}, Tom J and {Busko}, Ivo and {Donath}, Axel and {G{\"u}nther}, Hans Moritz and {Cara}, Mihai and {Lim}, P.~L. and {Me{\ss}linger}, Sebastian and {Burnett}, Zach and {Conseil}, Simon and {Droettboom}, Michael and {Bostroem}, Azalee and {Bray}, E.~M. and {Andersen Bratholm}, Lars and {Jamieson}, William and {Ginsburg}, Adam and {Barentsen}, Geert and {Craig}, Matt and {Pascual}, Sergio and {Rathi}, Shivangee and {Perrin}, Marshall and {Morris}, Brett M.},
        title = "{astropy/photutils: 2.2.0}",
         year = 2025,
        month = feb,
          eid = {10.5281/zenodo.14889440},
          doi = {10.5281/zenodo.14889440},
      version = {2.2.0},
    publisher = {Zenodo},
       adsurl = {https://ui.adsabs.harvard.edu/abs/2025zndo..14889440B}
}

@ARTICLE{Eastes2000C,
       author = {{Eastes}, Richard W.},
        title = "{Modeling the N$_{2}$ Lyman-Birge-Hopfield bands in the dayglow: Including radiative and collisional cascading between the singlet states}",
      journal = {\jgr},
         year = 2000,
        month = aug,
       volume = {105},
       number = {A8},
        pages = {18,557-18,573},
          doi = {10.1029/1999JA000378},
       adsurl = {https://ui.adsabs.harvard.edu/abs/2000JGR...10518557E}
}

@ARTICLE{Torr1994F,
       author = {{Torr}, Marsha R. and {Torr}, D.~G. and {Chang}, T. and {Richards}, P. and {Germany}, G.},
        title = "{N$_{2}$ Lyman-Birge-Hopfield dayglow from ATLAS 1}",
      journal = {\jgr},
         year = 1994,
        month = nov,
       volume = {99},
       number = {A11},
        pages = {21397-21408},
          doi = {10.1029/94JA01844},
       adsurl = {https://ui.adsabs.harvard.edu/abs/1994JGR....9921397T}
}

@ARTICLE{Cantrall2021A,
       author = {{Cantrall}, Clayton and {Matsuo}, Tomoko},
        title = "{Deriving column-integrated thermospheric temperature with the N$_{2}$ Lyman-Birge-Hopfield (2,0) band}",
      journal = {Atmospheric Measurement Techniques},
         year = 2021,
        month = nov,
       volume = {14},
       number = {11},
        pages = {6917-6928},
          doi = {10.5194/amt-14-6917-2021},
       adsurl = {https://ui.adsabs.harvard.edu/abs/2021AMT....14.6917C}
}

@ARTICLE{Hedin2009A,
       author = {{Hedin}, J. and {Gumbel}, J. and {Stegman}, J. and {Witt}, G.},
        title = "{Use of O$_{2}$ airglow for calibrating direct atomic oxygen measurements from sounding rockets}",
      journal = {Atmospheric Measurement Techniques},
         year = 2009,
        month = dec,
       volume = {2},
       number = {2},
        pages = {801-812},
          doi = {10.5194/amt-2-801-200910.5194/amtd-2-1419-2009},
       adsurl = {https://ui.adsabs.harvard.edu/abs/2009AMT.....2..801H}
}

@MISC{Ryon2019D,
       author = {{Ryon}, Jenna E. and {Avila}, Roberto J. and {Grogin}, Norman A. and {Bohlin}, Ralph},
        title = "{Bright Object Magnitude Limits for ACS/SBC and Color Corrections for All Three Channels}",
 howpublished = {Instrument Science Report ACS 2019-10, 15 pages},
         year = 2019,
        month = dec,
        pages = {10},
       adsurl = {https://ui.adsabs.harvard.edu/abs/2019acs..rept...10R}
}

@MISC{Guzman2024I,
       author = {{Guzman}, A.~M. and {Avila}, R.~J.},
        title = "{Updates to the SBC Dark Rate Monitor}",
 howpublished = {Instrument Science Report ACS 2024-04, 15 pages},
         year = 2024,
        month = sep,
        pages = {4-04},
       adsurl = {https://ui.adsabs.harvard.edu/abs/2024acs..rept....4G}
}

@MISC{Avila2016H,
       author = {{Avila}, R.~J. and {Chiaberge}, M.},
        title = "{Photometric Aperture Corrections for the ACS/SBC}",
 howpublished = {Instrument Science Report ACS 2016-5, 8 pages},
         year = 2016,
        month = sep,
        pages = {5},
       adsurl = {https://ui.adsabs.harvard.edu/abs/2016acs..rept....5A}
}

\end{document}